\documentclass[aps,pra,reprint,floatfix,superscriptaddress]{revtex4-2}

\usepackage{amsmath}
\usepackage{graphicx}
\graphicspath{{plots/}}
\usepackage{isotope}
\usepackage{physics}
\usepackage{nicefrac}
\usepackage{braket}
\usepackage{siunitx}
\begin{document}

\title[Article Title]{Measurement of the $\text{1}^\text{3}\text{S}_\text{1} \to \text{2}^\text{3}\text{S}_\text{1}$ interval in positronium using field-ionization of Rydberg states}

\author{Michael W. Heiss}
\email[Contact author: ]{michael.heiss@psi.ch}
\affiliation{Institute for Particle Physics and Astrophysics, ETH Z\"urich, CH-8093 Z\"urich, Switzerland}
\affiliation{Current address: PSI Center for Neutron and Muon Sciences CNM, 5232 Villigen PSI, Switzerland}

\author{Lucas de Sousa Borges}
\affiliation{Institute for Particle Physics and Astrophysics, ETH Z\"urich, CH-8093 Z\"urich, Switzerland}

\author{Artem Golovizin}
\affiliation{Institute for Particle Physics and Astrophysics, ETH Z\"urich, CH-8093 Z\"urich, Switzerland}
\affiliation{Current address: Lebedev Physical Institute LPI, 119991, Moscow, Russia}

\author{Paolo Crivelli}
\email[Contact author: ]{crivelli@phys.ethz.ch}
\affiliation{Institute for Particle Physics and Astrophysics, ETH Z\"urich, CH-8093 Z\"urich, Switzerland}

\date{\today}

\begin{abstract}
We report a new 40 ppb measurement of the positronium $\text{1}^\text{3}\text{S}_\text{1} \to \text{2}^\text{3}\text{S}_\text{1}$ interval using pulsed two-photon optical spectroscopy. The transition is detected via field-ionization of atoms excited from the 2S to the 20P Rydberg state. Precise Monte-Carlo line-shape simulations allow for the accounting of effects such as Doppler and AC Stark shifts, while an optical heterodyne measurement of the excitation laser pulse is used to correct for laser frequency chirp. A value of $1\,233\,607\,210.5\pm 49.6\, \mathrm{MHz}$ was obtained. This scheme allows for the measurement of the velocity distribution of positronium atoms to correct for the second-order Doppler effect. This is the major source of systematic uncertainty expected for future measurements of this transition with a CW laser, thus, our technique paves the way toward a new generation of a high precision determination of this interval in positronium. 
\end{abstract}

\maketitle

\section{\label{sec:introduction}Introduction}

Positronium is the lightest known atom and comprises a structureless point-like electron and its anti-particle, the positron \cite{Mohorovicic1934, Ruark1945}. As such, it is a purely leptonic system and can be described to very high precision by bound-state QED \cite{Karshenboim2004,ADKINS20221,Jentschura:2022xuc}, without the inherent complications given by the finite size of protonic atoms. Furthermore, being a true onium atom, recoil effects are strongly enhanced and its quantum numbers sum to zero, making it an ideal candidate to test fundamental symmetries like CPT-invariance \cite{Kostelecky2011,Kostelecky2015,Moskal:2021kxe} and for searches of new physics beyond the Standard Model \cite{ADKINS20221,Safronova2008, Frugiuele:2019drl, Vigo:2019bou}.

However, positronium is also a very challenging system for precision measurements due to its ephemeral nature. Being a bound state of anti-particles, it tends to self-annihilate quickly \cite{AlRamadhan1994,Kataoka2009} and due to its lightness it exhibits much larger velocities than other atoms at comparable temperatures. Nevertheless, being a precision test bench for QED with many unique features, it represents a particularly interesting system for spectroscopic measurements. 

With the advent of high power, narrow-band lasers at the required wavelength of $486\,\mathrm{nm}$, optical two-photon spectroscopy of positronium was realized by Chu and Mills measuring the $1^3\text{S}_1 \to 2^3\text{S}_1$ transition in 1982 \cite{Chu1982}. This was then improved upon, to be the benchmark QED test in positronium, reaching 12 ppb using pulsed lasers \cite{Chu1984} and a final precision of 3.2 ppb \cite{Fee1992,Fee1993,Fee1993a} using CW lasers. 

After nearly two decades without positronium laser experiments, the last ten years have seen a resurgence of precise measurements using lasers in positronium research (for an excellent overview see \cite{Cassidy2018}). However, the 1993 value for the $1^3\text{S}_1 \to 2^3\text{S}_1$ transition is still the most precise to date. Significant improvements in our understanding of bound state QED \cite{Czarnecki1999, Baker2014, Adkins2014, Eides2014, Adkins2014a, Eides2015, Adkins2015, Adkins2015a, Adkins2016, Eides2016, Eides2017, Adkins:2022coe, Eides2021, Eides2022, Eides2023} provide strong motivation for a new, more precise determination of this transition.

Here, we report a new measurement using multiple concurrent detection techniques. This includes a novel approach to further excite atoms from the 2S state to the 20P Rydberg state and subsequent field ionisation in a well-defined region of high electric field, allowing for a measurement of the atoms' time-of-flight and therefore their velocity distribution.

The experiment was conducted in two stages. Data for the first run was acquired without an absolute frequency standard and measurement of the laser chirp. For the second run, a Menlo FC1500-250-WG optical frequency comb and online laser chirp measurements were added to determine the absolute laser frequency accurately (see section \ref{sec:frequency-stabilization}). Additionally, the target and detector systems were redesigned for the second run, mainly to reduce the background due to backscattered positrons (see section \ref{sec:detection}). 

\section{\label{sec:experimental}Experimental methods}

\subsection{\label{subsec:beam}Slow positron beam and positronium production}

\begin{figure*}
\includegraphics[width=0.92\textwidth]{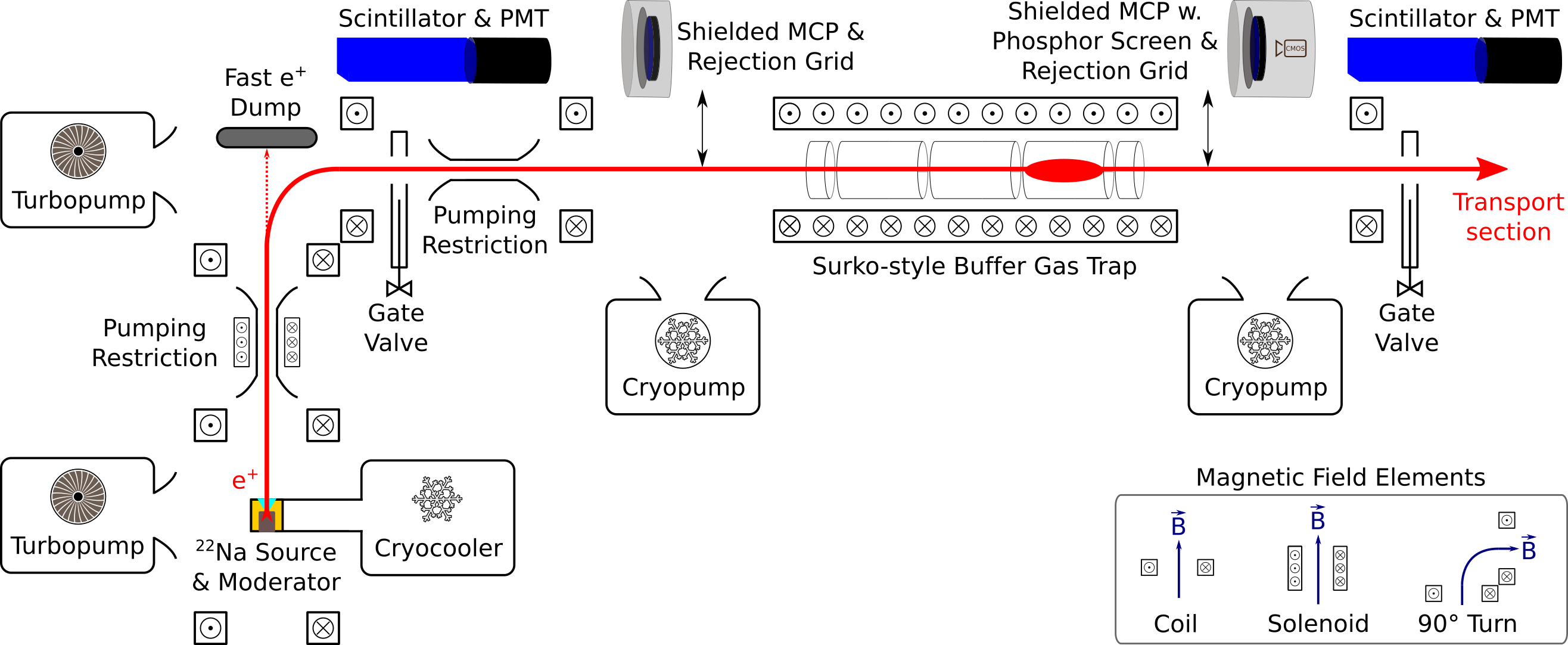} 
 \caption[Schematic overview of the pulsed slow positron beamline at ETH Zurich]{Schematic overview of the pulsed slow positron beamline at ETH Zurich. Positrons are created in the $\beta^+$ decay of the radioactive isotope $^{22}$Na and moderated using a solid rare gas moderator to approximately $40\,\mathrm{eV}$. They are then guided through a velocity selector to remove unmoderated positrons from the beam and transported to the buffer gas trap. There they lose energy by inelastic collisions with $\text{N}_2$ gas atoms to be trapped in an electric potential with magnetic radial confinement. In the last stage of the trap, the positron plasma is compressed by a rotating electric field and cooled by collisions with $\text{CF}_4$ gas. The positrons are then extracted in time-compressed bunches and guided to the experimental chamber.}
\label{fig:beam-schematic} 
\end{figure*}  

The pulsed slow positron beamline (see figure \ref{fig:beam-schematic}) provides less than $10\,\mathrm{ns}$ (FWHM) wide bunches of approximately $2 \times 10^4$ positrons per second at a repetition rate of 1-10 Hz, which are subsequently re-accelerated to 1-6 keV. Before entering the experimental chamber, they are extracted from the magnetic field and refocused spatially via an electrostatic einzel lens to approximately $\sigma_d = 1.5\,\mathrm{mm}$ beam diameter \cite{Cooke2015a, Heiss:2021qxv}. 

Positrons are then implanted on a mesoporous silica thin film target \cite{Liszkay2008}, where they can either be backscattered or penetrate the bulk and stop with a mean implantation depth on the order of a few hundred nanometers. A fraction of these positrons can form positronium by picking up an electron from the bulk material. Positronium rapidly thermalizes and is ejected into the pores with approximately $1\,\mathrm{eV}$ \cite{Nagashima1998} kinetic energy. The atoms will then diffuse through the porous network and rapidly cool down due to collisions with the pore walls \cite{Mariazzi2008}. This process will only continue as long as the de Broglie wavelength 
\begin{equation}
 \lambda_{dB} = \frac{h}{\sqrt{4 \pi m_e k_B T}} \approx \sqrt{\frac{0.24\,\mathrm{eV}}{E_{\mathrm{kin}}}} \cdot 1\,\mathrm{nm}
\end{equation}
is significantly smaller than the pore size of approximately $4\,\mathrm{nm}$ \cite{Crivelli2010}. As positronium approaches room temperature, the de Broglie wavelength becomes comparable to the pore size and the atom is confined quantum mechanically, which implies a significantly smaller diffusion coefficient \cite{Cassidy2010}. Experimental studies showed that the cooling from the classical to the quantum regime is completed within $5\,\mathrm{ns}$ and that the mean time for positronium to diffuse out of the sample depends linearly on implantation energy with approximately $3\,{\mathrm{ns}}/{\mathrm{keV}}$ \cite{Deller2015}.

Due to the quantum nature of positronium in the pores, the minimum energy of positronium cooling down in the network is not given by the thermal limit (approximately $30\,\mathrm{meV}$ for room temperature), but the ground state energy of the particle confined to the pore \cite{Crivelli2010}. For the simple model of a particle of mass $2m_e$ in an infinite spherical well potential, the ground state energy is given by \cite{Huang2016}:
\begin{equation}
 E_0 = \frac{h^2}{4 m_e d^2} \approx \left(\frac{27.4\,\mathrm{nm}}{d}\right)^2 \cdot 1\,\mathrm{meV} 
 \label{eq:gsenergy}
\end{equation}
which yields $E_0 \simeq 47\,\mathrm{meV}$ for $d = 4\,\mathrm{nm}$ pore size.

However, positronium emitted from these porous targets is not purely mono-energetic. The positrons have a broad implantation profile and, therefore, a fraction of positronium can form close to the surface. In this case, the atoms will be emitted epithermally before they cool down to the ground state energy of the pore. A distribution in pore sizes or positronium interacting on the target surface might also lead to smaller emission energies than the calculated ground state limit. In fact, while the exact velocity distribution is unknown, earlier experiments using the same type of target found that it could be well represented by a Maxwell-Boltzmann distribution with $T=900\,\mathrm{K}$ for $3\,\mathrm{keV}$ implantation energy \cite{Alonso2018,Deller2015}. Similarly, we found excellent agreement with a Maxwell-Boltzmann distribution for 700-550$\,\mathrm{K}$ at implantation energies of 4-5$\,\mathrm{keV}$. The most probable velocity of the distribution roughly corresponds to the velocity predicted by equation \ref{eq:gsenergy} and it is in good agreement with values found in earlier measurements for implantation energies above $4\,\mathrm{keV}$ \cite{Crivelli2010,Cassidy2010,Deller2015}. 

\subsection{\label{subsec:laser}Laser excitation}

\begin{figure*}
\includegraphics[width=0.92\textwidth]{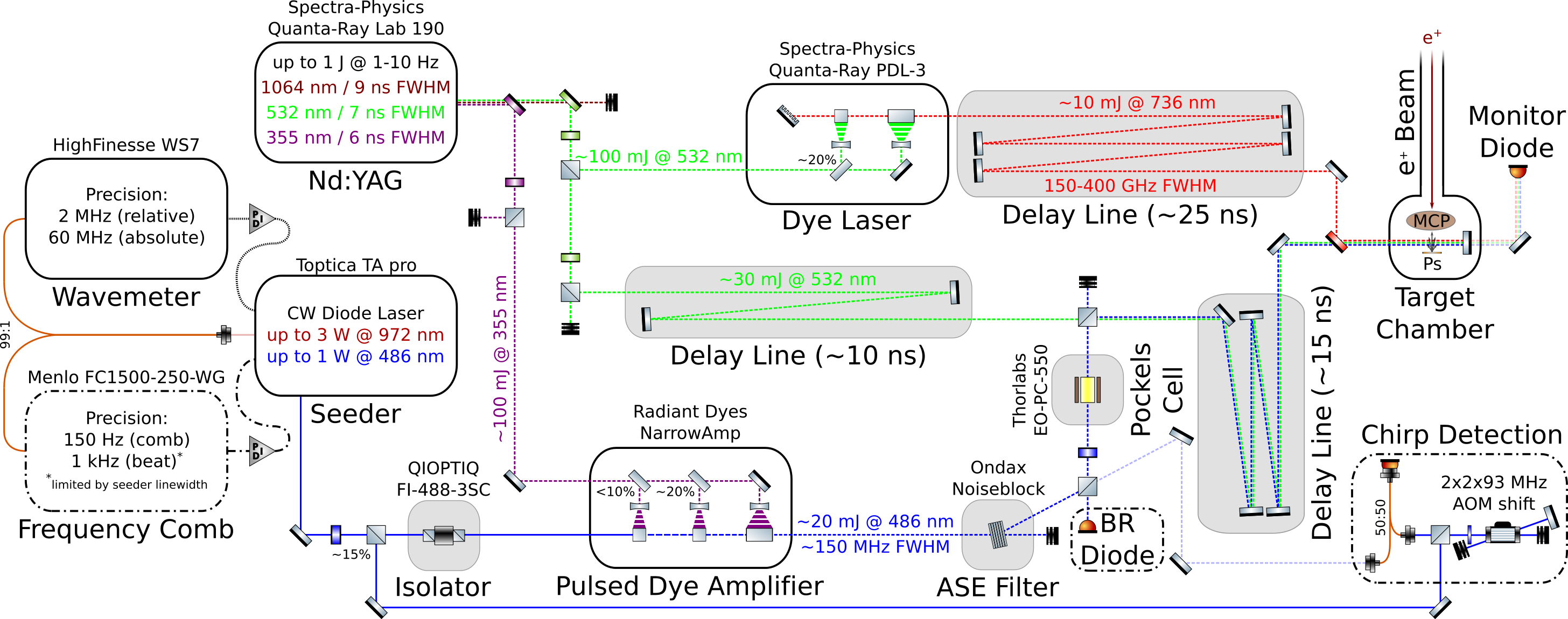} \\[0.15cm]
\includegraphics[width=0.69\textwidth]{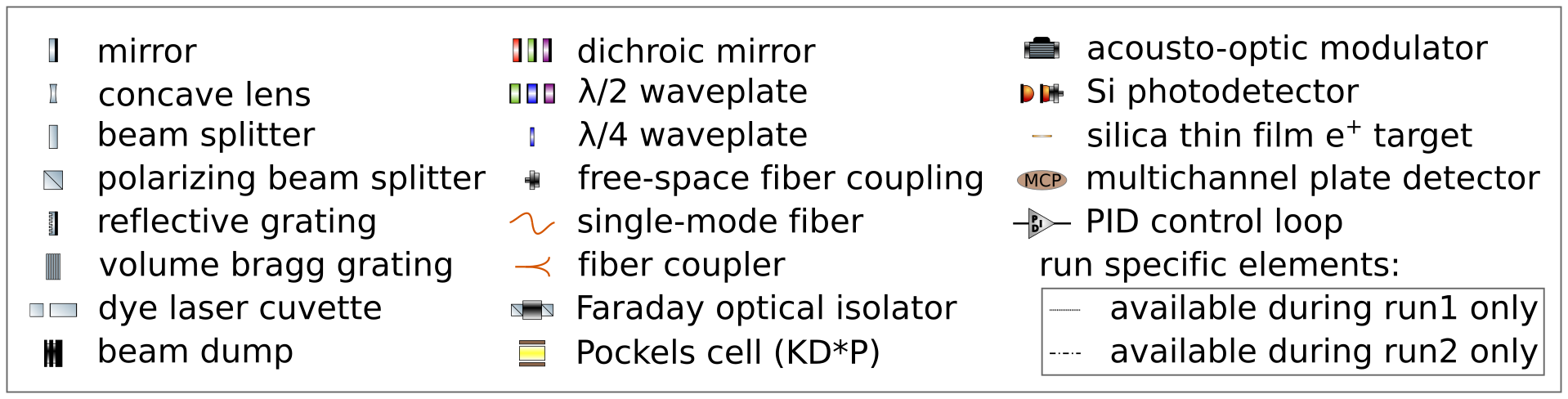} 
   \caption[Schematic overview of the 1S-2S laser excitation system for both runs.]{Schematic overview of the 1S-2S laser excitation system for both runs. Pulsed lasers are denoted as dotted and CW lasers as full lines. Not all elements are shown, e.g. pulse shaping optics, shutters etc. Positions, orientations and scales are arbitrary.   
\label{fig:laser-schematic}}
\end{figure*}

A schematic overview of the laser system employed to measure the 1S-2S transition is shown in figure \ref{fig:laser-schematic} (more details on the laser system can be found in \cite{Heiss:2021qxv}). Three separate laser beams at different wavelengths ($486\,\mathrm{nm}$, $532\,\mathrm{nm}$, and $736\,\mathrm{nm}$) are generated, combined and used to excite and photo-ionize positronium. A Spectra-Physics Quanta-Ray Lab 190 injection-seeded Nd:YAG allows for the generation of $532\,\mathrm{nm}$ (second harmonic) of up to $500\,\mathrm{mJ}$ and $355\,\mathrm{nm}$ (third harmonic) of up to $250\,\mathrm{mJ}$ with pulse lengths on the order of a few nanoseconds.

The excitation to the metastable $2\text{S}$ state is being driven by a modified Radiant-Dyes Pulsed Dye Amplifier (PDA), which is seeded at approximately $486\,\mathrm{nm}$ using a Toptica TA pro $972\,\mathrm{nm}$ diode laser and an included second harmonic generation stage, generating up to $1\,\mathrm{W}$ with a line-width on the order of $200\,\mathrm{kHz}$. The dye consists of Coumarin 102 \cite{Tuccio1973} along with a stabilizing agent (DABCO) \cite{VonTrebra1982} dissolved in absolute ethanol. It is pumped by the frequency-tripled output of the Nd:YAG laser, which is synchronized to the pulsed positron beam. Using a retro-reflecting mirror, this allows for two counter-propagating pulses of up to $20\,\mathrm{mJ}$ with a time spread of approximately $6\,\mathrm{ns}$ FWHM, centered approximately $3\,\mathrm{mm}$ away from the target. 

A Pockels cell is used to isolate the PDA from the pulse returning from the retro-reflection mirror in the experimental chamber. This also allows to vary the fraction of the blue pulse transmitted to the chamber by changing the peak-to-peak value of the voltage supplied to its electrodes. By continuously measuring the power transmitted to the experimental chamber long-term stabilization of the average pulse energy (e.g. to counteract dye degradation) to better than $1\,\%$ is achieved. The addition of a diode on one of the beamsplitters allows to limit the maximal beam misalignment between the incoming and retro-reflected beam axes to be below $1.4\,\mathrm{mrad}$ for the beam path length of approximately $5\,\mathrm{m}$.

Since the photo-ionization cross-section is around a factor 3 higher than the excitation cross-section, a significant fraction -- on the order of a few percent, when optimized for efficient excitation -- of positronium is ionized in the same laser by a third photon. The freed positrons can be detected, either by their annihilation signal in a scintillator \cite{Cooke2015} or by measuring them directly in an MCP placed approximately $4\,\mathrm{cm}$ from the target. To guide the positrons from photo-ionization to the detector, a bias potential on the order of $100\,\mathrm{V}$ is applied to the target holder, resulting in $\lesssim 10\,\mathrm{V}/\mathrm{cm}$ electric field in the laser excitation region.

While direct photo-ionization in the exciting laser can be used for detection, there is significant broadening and shift due to the AC-Stark effect. Furthermore, this does not allow for direct detection of the surviving 2S states. Using an additional laser with photon energies higher than the binding energy $1.7\,\mathrm{eV}$ of the 2S (or any other excited) state allows for delayed photo-ionization independent of the 1S-2S exciting laser. For this purpose, approximately $30\,\mathrm{mJ}$ of the frequency-doubled $532\,\mathrm{nm}$ output of the Nd:YAG and a delay line on the order of $10\,\mathrm{ns}$ were used. This ionizes any excited state positronium with a probability close to unity. 

Alternatively, the excited state positronium atoms can further be excited to a Rydberg state, using a modified Spectra-Physics Quanta-Ray PDL-3 pulsed dye laser pumped by the second harmonic of the Nd:YAG. The 20P state was selected as a compromise between polarizability, fluorescence lifetime and losses due to collisions with residual gas atoms in our vacuum environment, which was operated at approximately $10^{-8}\,\textrm{mbar}$ of residual pressure (see appendix \ref{sec:appendix-tau}). Pulses at $736\,\mathrm{nm}$ of energies up to $10\,\mathrm{mJ}$ at approximately $7\,\mathrm{ns}$ FWHM were produced using Styryl 7 dye \cite{Kato1980}. Since the transition $2\text{S} \to 20\text{P}$ is a one-photon process, the first-order Doppler shift due to the distribution of the emission angles had to be covered by enlarging the laser bandwidth to values on the order of $300\,\mathrm{GHz}$ (FWHM).

The wavelength and bandwidth of the output of the PDL-3 dye laser are regularly measured using a calibrated ASEQ Instruments HR1 spectrometer with $50\,\mathrm{GHz}$ resolution and show no significant drift. The drive system for the grating used in the frequency selection of the PDL-3 system allows for changes with a repeatability of better than $5\,\mathrm{GHz}$ and frequency scans over a range of many nanometers.

\subsection{Frequency stabilization and measurement\label{sec:frequency-stabilization}} 

A HighFinesse WS7 wavemeter is used to monitor the output of the Toptica TA pro diode laser used for seeding the PDA, with an absolute precision of $60\,\mathrm{MHz}$ \footnote{It should be noted, that the internal calibration lamp of the device was not functioning at the time of data-taking and the device specification states that the absolute error is guaranteed only if the device is frequently calibrated. However, operating the device in conjunction with a frequency comb over several months, we did not measure any shifts above approximately $30\,\mathrm{MHz}$ from the stored calibration value.}. For the first run the internal PID control of this device was used to stabilize and scan the frequency of the seed laser.

For the second run, a Menlo FC1500-250-WG optical frequency comb was used as an absolute frequency standard, which was referenced to a precision GPS clock. In addition to measuring the frequency, the beatnote output is used to lock the seed laser to a comb tooth \cite{Schuenemann1999}. Not only does this counter possible drift, it also allows to scan over several $100\,\mathrm{MHz}$ by changing the repetition rate of the frequency comb. 

\begin{figure}
\includegraphics[width=\columnwidth]{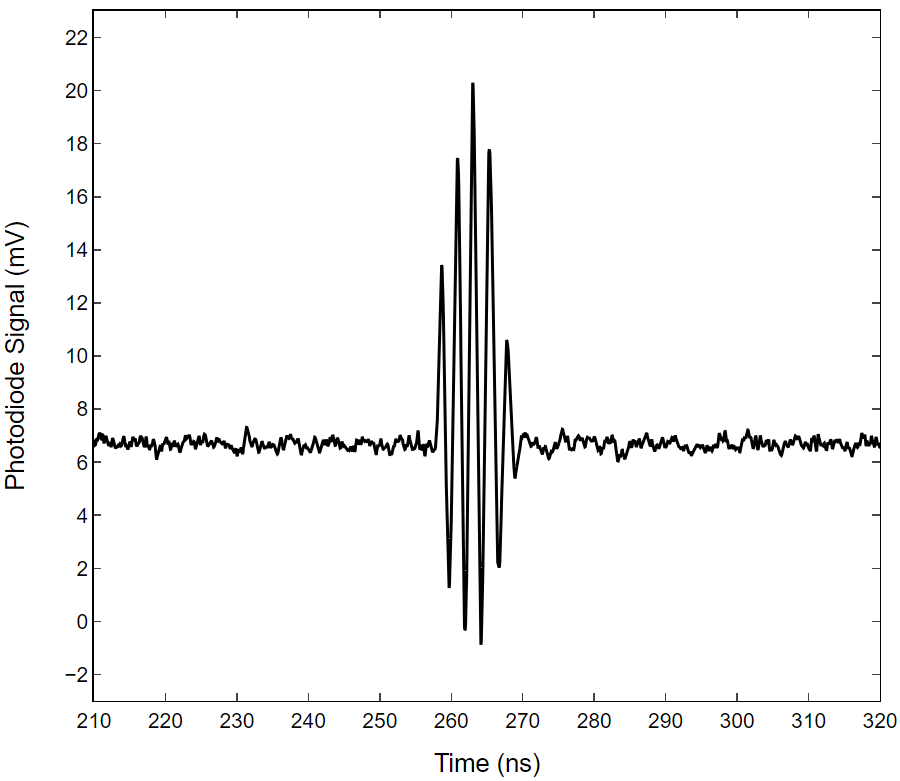} 
   \caption[Typical chirp beatnote oscilloscope waveform taken during run 2.]{Typical chirp beatnote oscilloscope waveform taken during run 2.
\label{fig:chirp-single}}      
\end{figure}

Frequency chirp \cite{Wieman1980,Danzmann1989} shifts the output of the pulsed dye amplifier from the frequency of the seed light due to the rapid change in the refractive index of the gain medium \cite{Reinhard1996}. An optical heterodyne technique to measure and correct for this shift \cite{Chu1982,Chu1984,Fee1992} pulse-by-pulse was implemented for run 2. Part of the seed light is shifted in frequency using an AOM in a second order double pass configuration operating at $92.96\,\mathrm{MHz}$ and mixed with the output of the PDA using a fiber-optic coupler. 
The obtained high beatnote frequency at near $370\,\mathrm{MHz}$ is needed for precisely measuring the frequency chirp of a few ns wide pulses. The resulting beatnote is then recorded using a fast Thorlabs DET025AFC photo-diode and a $1\,\mathrm{GHz}$ bandwidth Teledyne LeCroy Waverunner 8104 oscilloscope at $20\,\mathrm{GS/s}$ sampling rate (see Figure \ref{fig:chirp-single}). 

\begin{figure}
\includegraphics[width=\columnwidth]{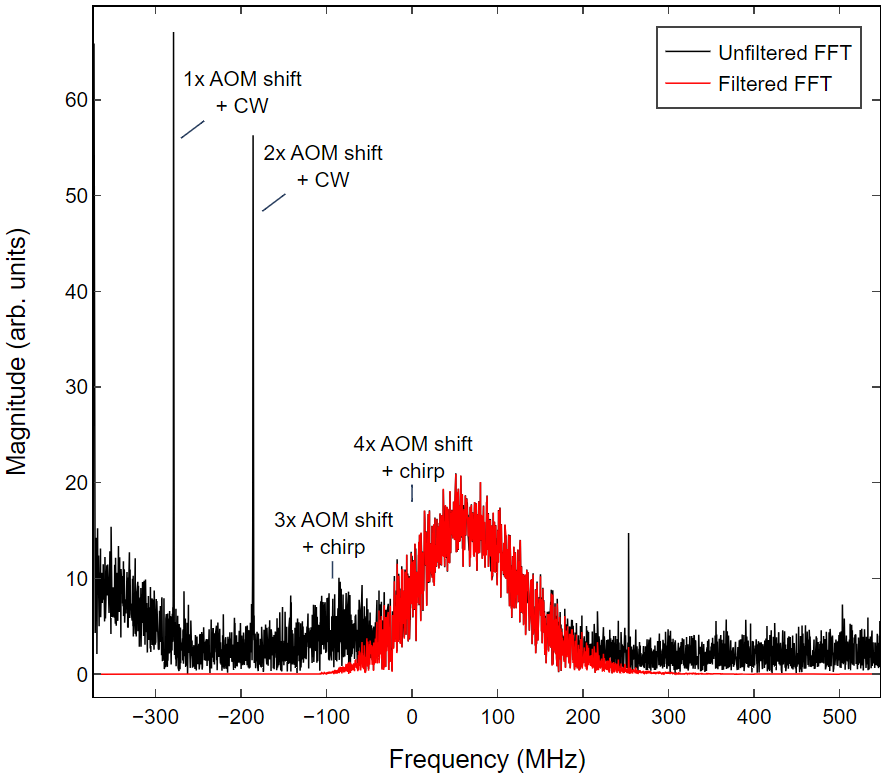} 
   \caption[One-sided FFT spectrum (after subtraction of total AOM shift) of a typical chirp beatnote taken during run 2.]{One-sided FFT spectrum (after subtraction of total AOM shift) of a typical chirp beatnote taken during run 2 before (black) and after (red/grey) filtering.
\label{fig:chirp-spectrum}}
\end{figure}

To extract the instantaneous frequency of the beatnote (which corresponds to the AOM shift plus laser chirp) the signal must first be filtered (see figure \ref{fig:chirp-spectrum}). This is necessary to remove high frequency noise, components related to shifts other than twice the second order of the AOM and to the pulse envelope. To avoid any filter-induced delay, a zero-phase Butterworth band-pass software filter \cite{Oppenheim1998,Gustafsson1996} is implemented in the analysis. This ensures that the chirp calculated from the filtered signal does indeed correspond to the specific timing of the pulse.

\begin{figure}
\includegraphics[width=\columnwidth]{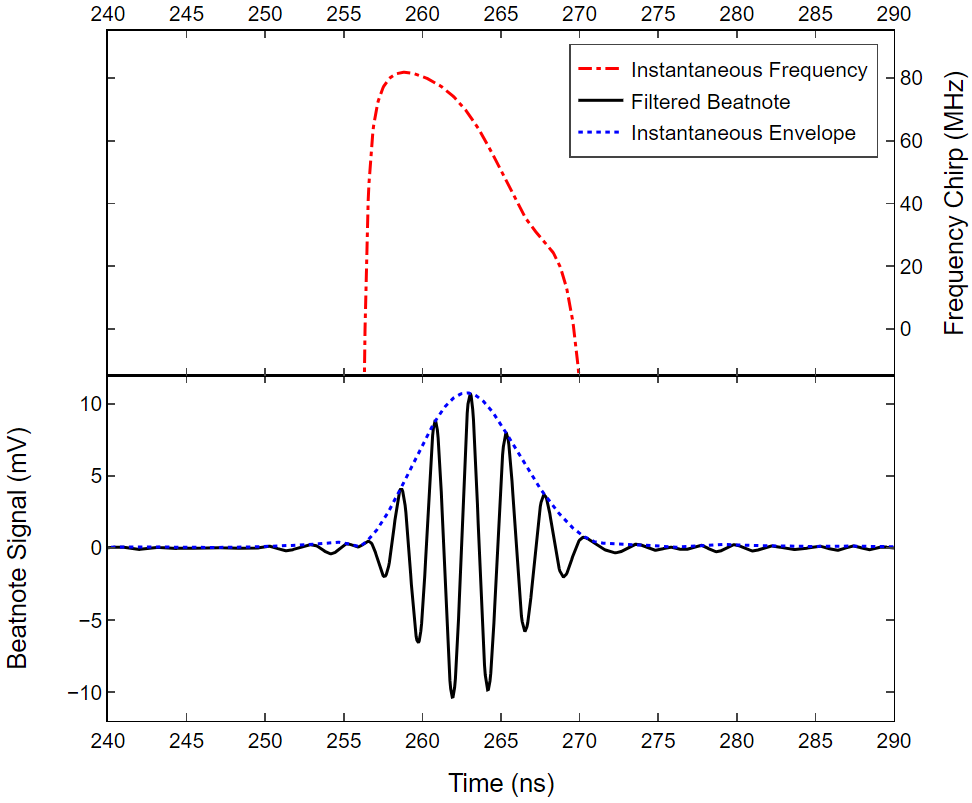} 
   \caption[Filtered beatnote, instantaneous envelope and frequency, obtained via Hilbert transformation.]{Filtered beatnote (solid black), instantaneous envelope (dashed blue) and frequency (dashdotted red), obtained via Hilbert transformation. Note that the instantaneous frequency is only well defined for times with appreciable pulsed laser intensity, which is why it is only shown here for intensities of at least $1\%$ of the peak value.
\label{fig:chirp-final}}       
\end{figure}

After filtering, a Hilbert transformation \cite{Luo2009} is used to extract both the instantaneous envelope and frequency (see figure \ref{fig:chirp-final}). The chirp correction for a specific beatnote is obtained by averaging the instantaneous frequency weighted by the intensity profile. This is done on a pulse-by-pulse basis and each event can then be assigned the chirp corrected frequency. 

The calculated value of this correction depends on the specific choice of cut-on and cut-off frequencies. Varying these parameters in our analysis, we found that this introduces a systematic error on the order of $1\,\mathrm{MHz}$ on the determination of the chirp corrected frequency.

\subsection{\label{sec:detection}Detection}

\begin{figure}
\includegraphics[width=0.9\columnwidth]{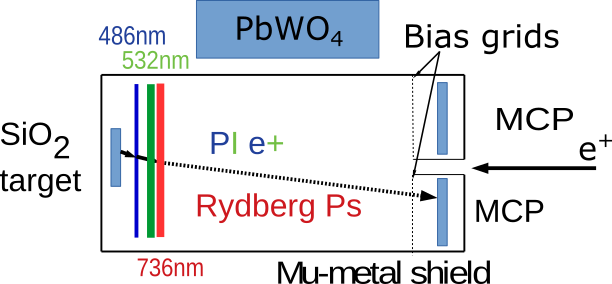}%
\caption[Sketch of positronium excitation and detection scheme]{Sketch of positronium excitation and detection scheme: Positrons from the bunched positron beam are focused onto the silica thin film target. There it is converted to positronium and emitted into vacuum. It is then laser excited to the 2S state by the $486 \, \mathrm{nm}$ laser and subsequently either ionized by a $532 \, \mathrm{nm}$ laser or further excited to the 20P state using a $736 \, \mathrm{nm}$ laser. The positrons released either from (direct and/or delayed) photo-ionization or from field-ionization of the Rydberg atoms between the grids and the MCP, can be directly detected on the MCP and via their annihilation photons in the PbWO$_4$ scintillator.
\label{fig:detection-scheme}}
\end{figure}

A schematic overview of the detection can be found in figure \ref{fig:detection-scheme}. The implantation of positrons is monitored by the prompt peak in a lead tungstate scintillator and by measuring backscattered positrons in an MCP that is placed at approximately $4\,\mathrm{cm}$ distance to the target. Positrons from direct photo-ionization in the exciting laser as well as from delayed photo-ionization are guided by a $\simeq 10\,\mathrm{V}/\mathrm{cm}$ electric field between the target and the MCP detector assembly.

Positronium atoms in Rydberg states traverse the distance between excitation and detector relatively unperturbed and only ionize in the strong electric field between a grounded grid and the MCP front plate of approximately $4\,{\mathrm{kV}}/{\mathrm{cm}}$ after a time of flight of a few hundred nanoseconds. This allows for the additional measurement of the velocity distribution of positronium atoms, which can be used to measure the second order Doppler shift. The addition of a position-sensitive detector or parabolic grids to field-ionize the atoms in well-defined positions can be used in future experiments to correct this major systematic on a per-atom basis.

\begin{figure}
\includegraphics[width=\columnwidth]{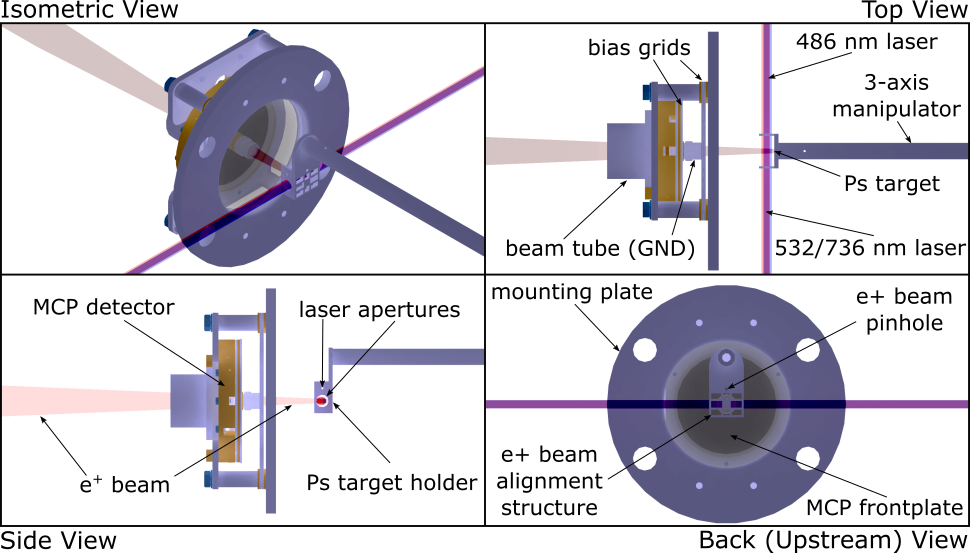}
\caption[MCP detector and Ps conversion target assembly for run 1.]{MCP detector and Ps conversion target assembly for run 1. Most parts of the mounting structure, magnetic shielding and vacuum chamber are not shown.
\label{fig:Run1_MCPs}}
\end{figure}

Figure \ref{fig:Run1_MCPs} shows the detector and target assembly used during run 1. A matched set of two center-hole MCP plates is used in a chevron configuration, mounted in a custom assembly. A grounded tube through the center axis of the detector allows the positron beam to pass, which results in the active area being an annulus with $9.14\,\mathrm{mm}$ ID and $40\,\mathrm{mm}$ OD. The assembly incorporates two independent bias grids ($92\,\%$ transmission), placed in approximately $3\,\mathrm{mm}$ and $13.5\,\mathrm{mm}$ distance from the the front plate, respectively. For this measurement both grids were set to ground potential. The center of the $486\,\mathrm{nm}$ laser axis and the center of the grid on which the 20P states field ionize are separated by $\left(38.0\pm1.0\right)\,\mathrm{mm}$. The kinetic energy of positrons implanted on the porous silica thin film target was $4\,\mathrm{keV}$ for this run.

\begin{figure}
\includegraphics[width=\columnwidth]{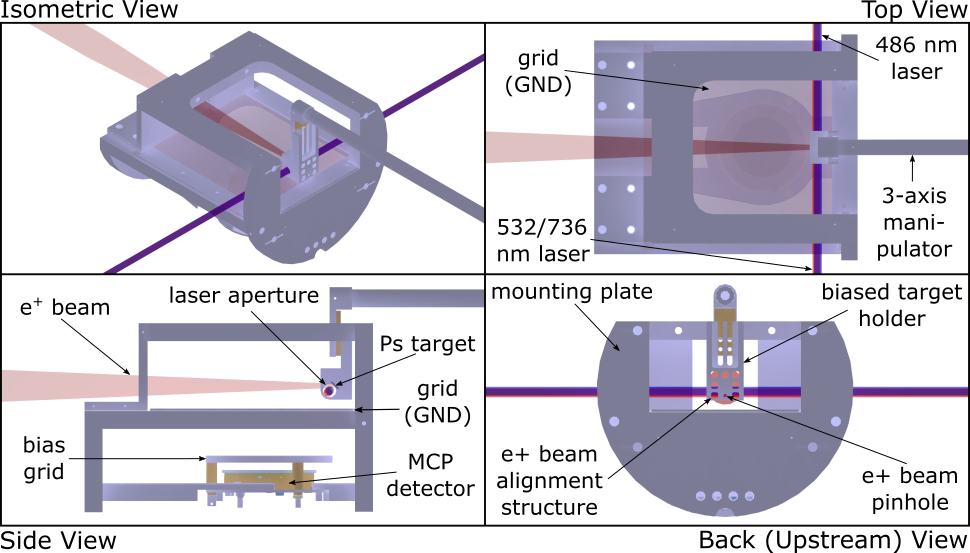}
\caption[MCP detector and Ps conversion target assembly for run 2.]{MCP detector and Ps conversion target assembly for run 2. Most parts of the mounting structure, magnetic shielding and vacuum chamber are not shown.
\label{fig:Run2_MCPs}}
\end{figure}

For run 2 the detector geometry was modified to reduce background from backscattered positrons, as well as to reduce losses of approximately $20\,\%$ through the central aperture (see figure \ref{fig:Run2_MCPs}). To this end, positrons are implanted at a $45^\circ$ angle of incidence, which necessitates for the kinetic energy of the impinging positrons to be increased to achieve the same depth profile as for $0^\circ$ angle of incidence. The implantation energy was chosen to be $6\,\mathrm{keV}$, which corresponds to implanting with approximately $5\,\mathrm{keV}$ perpendicular to the target.

The MCP used two matched plates with $25\,\mathrm{mm}$ quality diameter in a chevron configuration, mounted in a custom assembly. A bias grid is mounted at approximately $6\,\mathrm{mm}$ distance to the front plate of the MCP, while a second grid is placed on the mounting structure in $27.5\,\mathrm{mm}$ distance. Both tungsten wire mesh grids, allowing for approximately $92\,\%$ transmission, were grounded for this measurement. However, due to the grid plane orientation of $45^\circ$ to the target, an additional loss factor of ${1}/{\sqrt{2}}$ per grid is expected on average. The center of the $486\,\mathrm{nm}$ laser axis and the center of the grid, on which the 20P states field ionize, are separated by $\left(41.8\pm0.5\right)\,\mathrm{mm}$.

\begin{figure}
\includegraphics[width=\columnwidth]{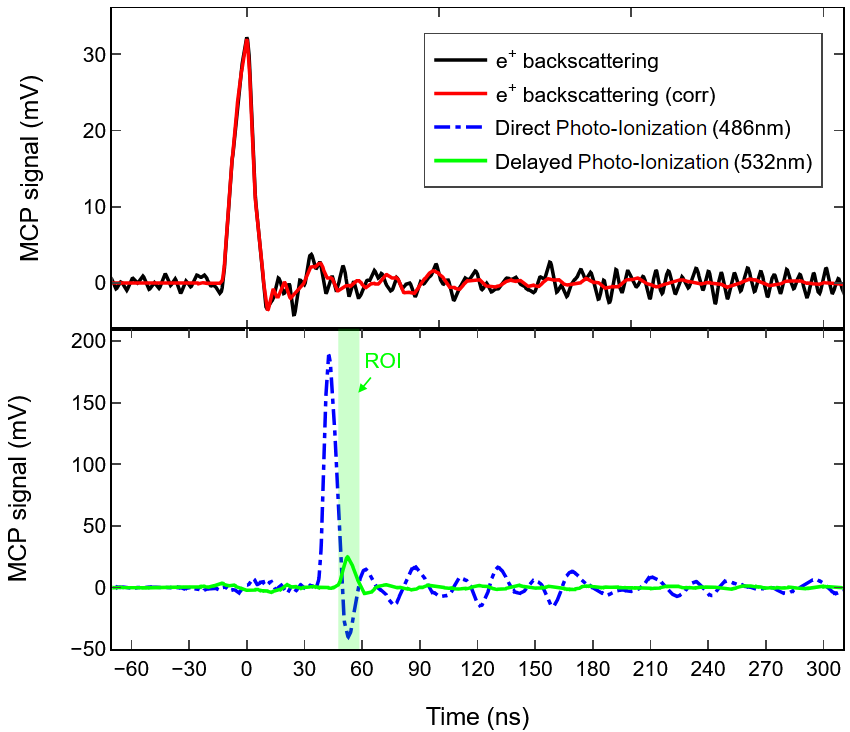}
\caption[Averaged MCP waveforms taken during run 1.]{Averaged MCP waveforms taken during run 1. Above for backscattered positrons without laser excitation before (black) and after (red/grey) HV switching noise correction. Below for direct photo-ionization (blue dashdotted) after subtraction of backscattered positrons and delayed photo-ionization (green solid) after subtraction of direct photo-ionization signals. Also shown below as a shaded region is the ROI for the integration of the delayed photo-ionization signal.
\label{fig:MCP-compare}}
\end{figure}

\begin{figure}
\includegraphics[width=\columnwidth]{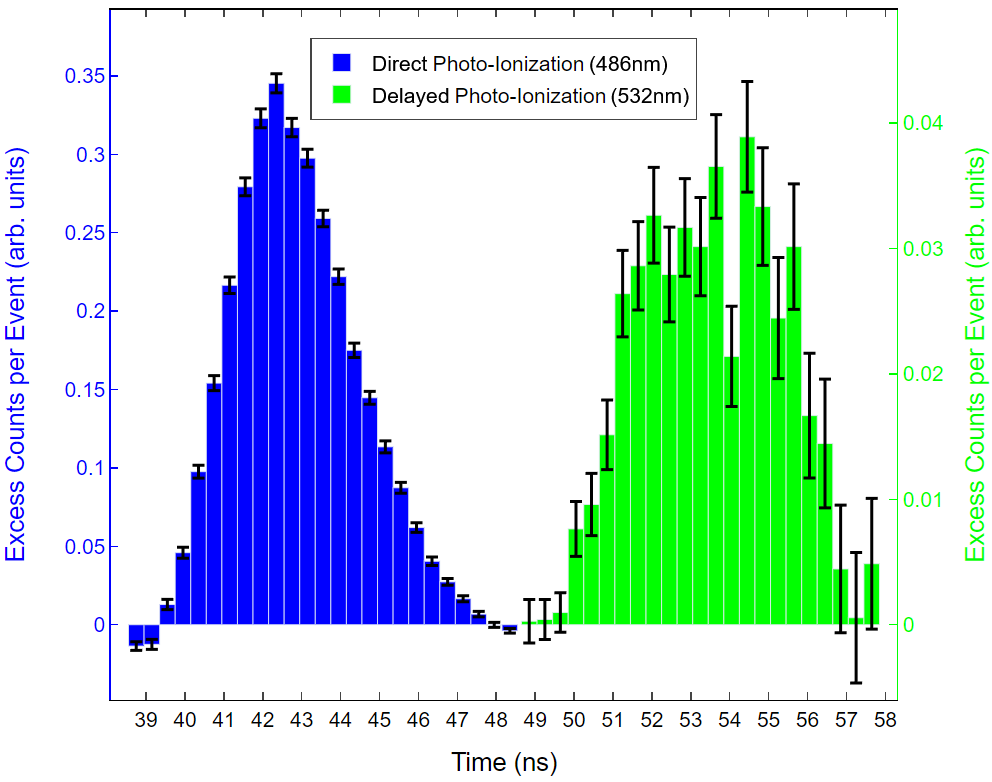}
\caption[Histogram of excess MCP counts due to photo-ionization.]{Histogram of excess MCP counts due to photo-ionization taken during run 1 (direct photo-ionization in blue/dark-grey and delayed photo-ionization in green/light-grey, respectively). The time axis is relative to the implantation of positrons on the porous silica target.
\label{fig:Hist_PI_run1}}
\end{figure}

\begin{figure}
\includegraphics[width=\columnwidth]{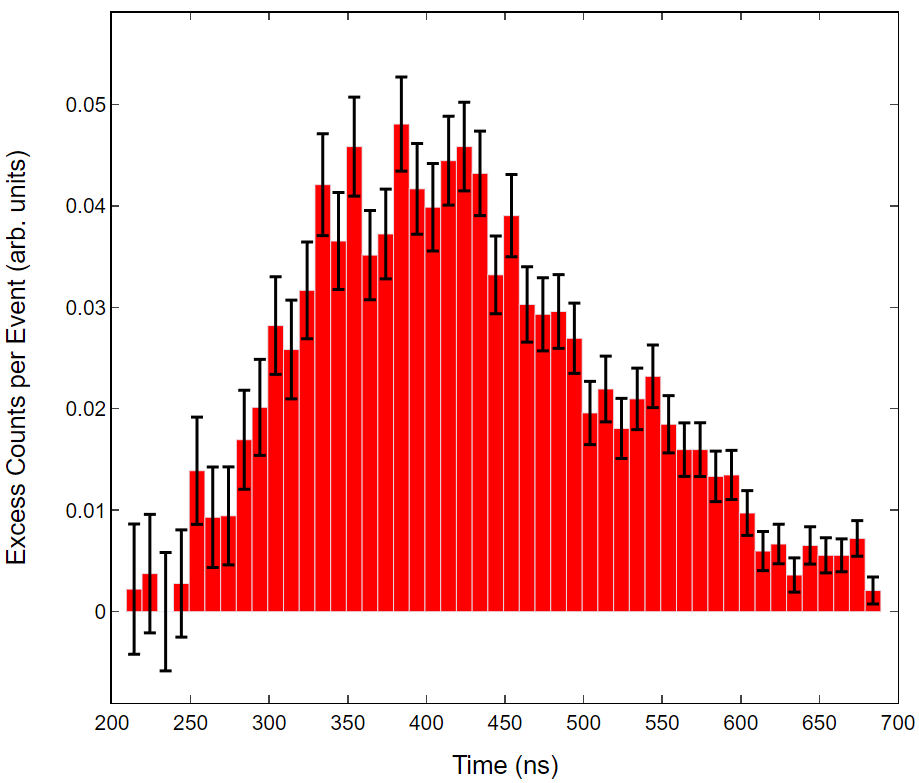}
\caption[Histogram of excess MCP counts due to Rydberg field ionization.]{Histogram of excess MCP counts due to Rydberg field ionization taken during run 1. The time axis is relative to the implantation of positrons on the porous silica target.
\label{fig:Hist_FI_run1}}
\end{figure}

Since most events of interest are due to a single positron and multiple overlapped counts are rare, individual peaks can be counted after the respective backgrounds are subtracted (see Figure \ref{fig:MCP-compare}) and filtered. Figures \ref{fig:Hist_PI_run1} and \ref{fig:Hist_FI_run1} show the time spectra of excess counts for photo-ionization and field ionization for the dataset of run 1. 

\section{\label{sec:results}Results and Discussion}

\begin{figure}
\includegraphics[width=\columnwidth]{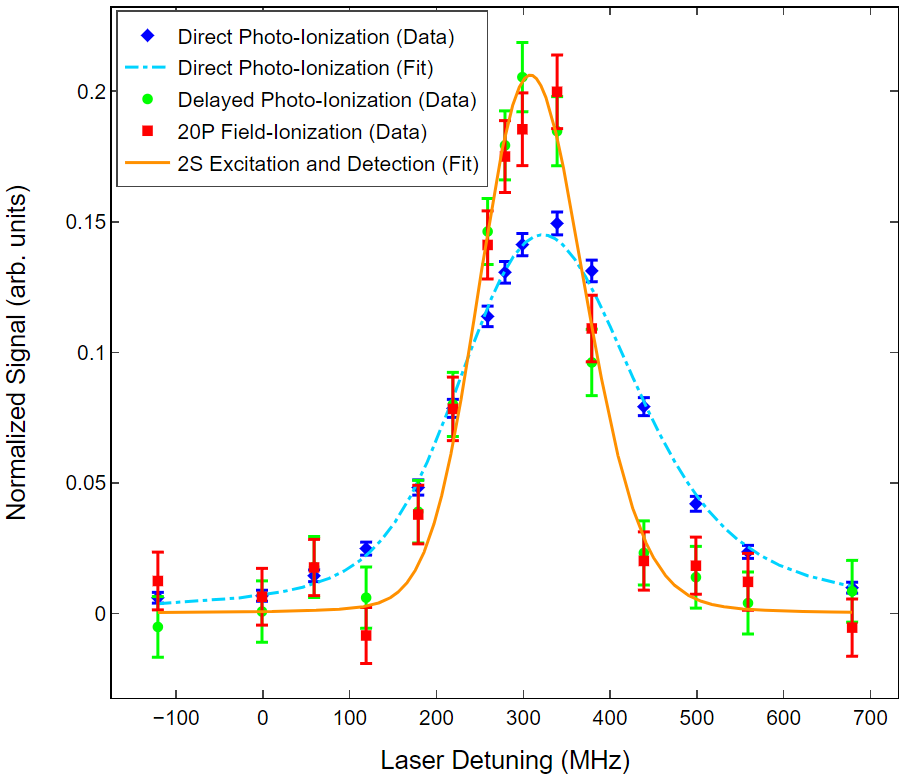} 
  \caption[Positronium direct photo-ionization and 2S excitation measured via delayed photo-ionization and field-ionization as a function of laser detuning for data taken during run 1.]{Positronium direct photo-ionization (blue diamonds) and 2S excitation measured via delayed photo-ionization (green circles) and field-ionization (red squares) as a function of laser detuning for data taken during run 1. Note that the laser detuning is given as set on the wavemeter and the systematic error is not shown. Furthermore, the chirp was not measured for this dataset and is therefore not corrected here. The data is sum-normalized and the best fits for all simulated parameter sets for run 1 is shown (light-blue dashdotted line for direct photo-ionization and orange solid line for 2S excitation).
\label{fig:analysis-line-fit}}
\end{figure}

\paragraph{Results of run 1:} Figure \ref{fig:analysis-line-fit} shows the sum-normalized measured data-points taken during run 1. The results for direct photo-ionization and field-ionization were extracted by directly counting events (peaks above threshold), while for delayed photo-ionization the average MCP signal was integrated in the region of interest (see figure \ref{fig:MCP-compare}). Due to the relatively short time delay of approximately $10\,\mathrm{ns}$ between direct and delayed photo-ionization, reliably identifying peaks was severely impeded by ringing associated with both ion feedback and imperfect impedance matching of the MCP. The frequency shown corresponds to the value set via PID control on the wavemeter and the systematic error is omitted in the graph. 

The process of extracting the transition frequency from these results proceeds in three steps. The line-shape calculated by a detailed Monte-Carlo simulation (see Appendix \ref{sec:lasersim-desc}) is fitted simultaneously to all normalized data-points with the only free parameter, aside from the amplitude scaling, being a common offset from zero laser detuning. This allows to establish a correspondence of measured to simulated frequencies for this specific set of parameters. 
 
\begin{figure}
\includegraphics[width=\columnwidth]{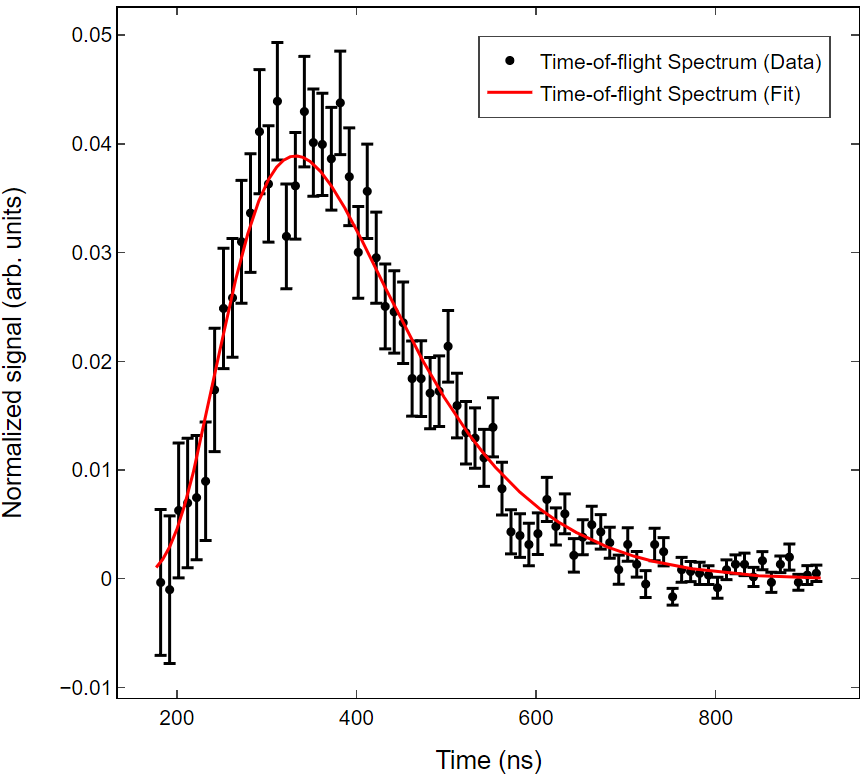} 
   \caption[Normalized time-of-flight spectrum and best fit for all simulated parameter sets for run 1.]{Normalized time-of-flight spectrum relative to the timing of the 20P excitation to field-ionization of the Rydberg atoms on the MCP and the corresponding best fit for all simulated parameter sets for run 1.
\label{fig:analysis-tof-fit}}
\end{figure}

To achieve the best signal-to-noise ratio for the time-of-flight spectrum, a subset of frequency points around the peak of the transition is chosen to build the histogram. However, due to the Doppler shift the exact shape of the time-of-flight spectrum is unique for each frequency, since certain velocity components are favored for certain offsets. Therefore, the frequency correspondence established in step 1 of the fitting process allows for building the matching histogram from simulation. This histogram is then fitted with a $\chi$-distribution as outlined in Appendix \ref{subsec:simlaser} for each set of parameters available (see figure \ref{fig:sim-fit-tof}). Note that the simulated time-of-flight distribution assumes a Maxwell-Boltzmann distribution for the initial positronium velocity and the data shows excellent agreement. However, as mentioned above, it should be reiterated that there is no reason to expect positronium to be intrinsically thermal at these energies. 

Finally, the third step in the fitting procedure entails repeating the process of step one, additionally fitting the normalized time-of-flight spectrum obtained in the second step. The only free parameter in this additional fitting step is the overall amplitude. Figure \ref{fig:analysis-tof-fit} shows the fitted function of the best parameter set for the respective time-of-flight spectrum. This combined fitting procedure allows for generating a single $\chi^2$ value, which is used to find the best fit for the complete set of parameters simulated.

\begin{figure}
\includegraphics[width=\columnwidth]{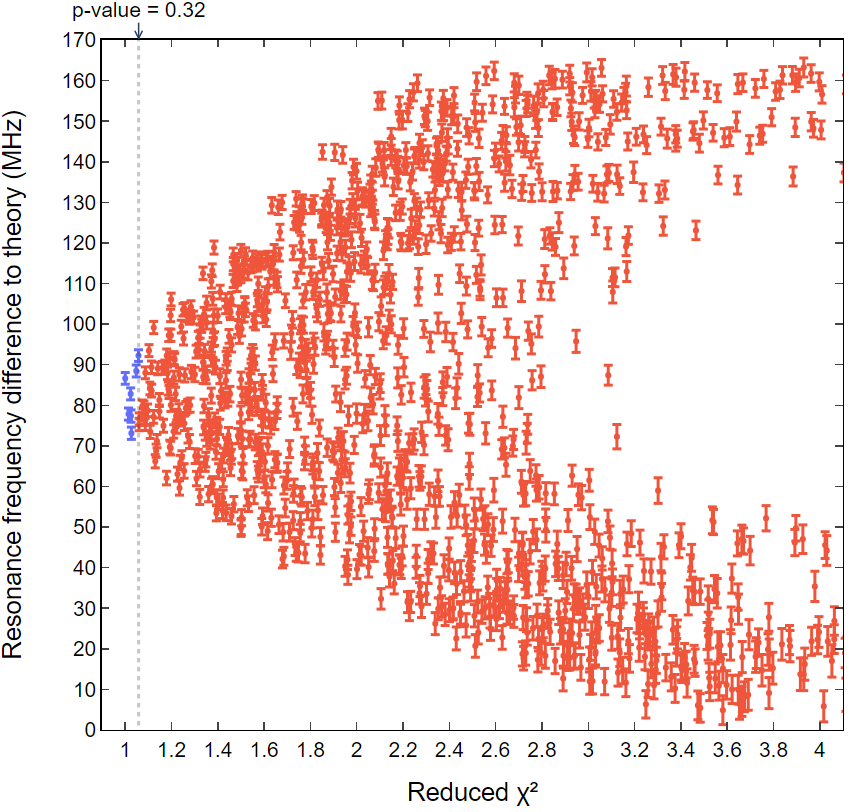} 
   \caption[Plot of reduced ${\chi}^2$ values for the fits of the dataset of run 1 to simulated line-shapes and TOF spectra.]{Plot of reduced ${\chi}^2$ values for the fits of the dataset of run 1 to simulated line-shapes and TOF spectra. The cut-off for a p-value of 0.32 is shown as a dotted line. Fits below correspond to be within a $68\%$ confidence interval and are taken into account for the estimation of the systematic error.
\label{fig:chi2-minimization}}
\end{figure}

Aside from determining the best fitting set of parameters, this method allows for the extraction of an estimate for the systematic error as well. For each individual fit a $\chi^2$ test determines the quality-of-fit \cite{Bevington2003} to the simulated line-shape and time-of-flight spectrum. The reduced $\chi^2$ parameter 
\begin{equation}
 {\chi_r}^2 = \frac{{\chi}^2}{\text{d.o.f.}} = \frac{1}{\text{d.o.f.}} \sum_i \frac{ (O_i - F_i)^2}{ \sigma_i^2 }
\end{equation}
is computed, where $O_i$ and $F_i$ correspond the the i-th observed and fit value, respectively. A cut-off on ${\chi_r}^2$ for a p-value of 0.32 then corresponds to selecting all simulated zero-detuning frequency values that fit the dataset with a confidence level of $68\,\%$, which we will use as an estimator for the systematic error. However, since the true variance $\sigma_i^2$ for each data point is not known a priori, this method can lead to underestimating the systematic error, due to partly unaccounted variance of the dataset or an effect that was not accounted for in the simulation. Therefore, in the case of $\text{min}({\chi_r}^2) > 1$, the dataset is corrected by multiplying an overall error scaling factor and repeating the fitting procedure, such that $\text{min}({\chi_r}^2) = 1$.

\begin{table} [ht!]
  \begin{tabular}
    {| c || c |}
    \hline
    Magnitude & Description \\
    \hline
    \hline
    $1.4\,\mathrm{MHz}$ & Data Fitting Error  \\
    $0.9\,\mathrm{MHz}$ & Simulation Fitting Error  \\
    $9.6\,\mathrm{MHz}$ & Systematic Fitting Error  \\
    $15.0\,\mathrm{MHz}$ & Chirp Estimation Error \\
    $60.0\,\mathrm{MHz}$ & Absolute Frequency Reference Error \\
    \hline
    $62.6\,\mathrm{MHz}$ & Quadrature Sum \\
    \hline
   \end{tabular} 
\caption{Sources and magnitude of statistical and systematic errors present in result for run 1.
\label{tab:errors_run1}}
\end{table}

Correcting for the chirp in the result is necessary, since our measurements showed frequency chirp of approximately $60\,\mathrm{MHz}$ (see discussion of run 2 results below) for this laser system operated in a similar regime as in run 1. The frequencies were therefore corrected by $\Delta \nu_c = (60 \pm 15)\,\mathrm{MHz}$, where the error was conservatively estimated by varying operational parameters within reasonable limits, while measuring the respective laser chirp.

The final result for the resonance frequency in terms of laser detuning obtained during run 1 reads
\begin{equation}
 616\,803\,638\:\mathrm{MHz} \pm 63\:\mathrm{MHz}
\end{equation}
where the respective sources of error can be found in table \ref{tab:errors_run1}.
 
This corresponds to a transition frequency of
\begin{equation}
 1\,233\,607\,275\:\mathrm{MHz} \pm 125\:\mathrm{MHz}
\end{equation}
or equivalently, this result differs from theory by
\begin{equation}
 53\,\mathrm{MHz} \pm 125\,\mathrm{MHz} \pm 0.6\,\mathrm{MHz} 
\end{equation}
where $0.6\,\mathrm{MHz}$ corresponds to the theoretical uncertainty in the 1S-2S interval at $m\alpha^6$.

\paragraph{Sources of systematic error and respective magnitudes:} To ascertain the relative impact of different sources to the systematic error estimate of run 1, a set of additional simulations were produced, varying only individual parameters -- in the case of misalignment of the incoming and retro-reflected laser beam, both angles in the XY and XZ plane were scanned. The systematic error corresponding to the respective parameter was then estimated analogously to the method described above. 

\begin{figure}
\includegraphics[width=\columnwidth]{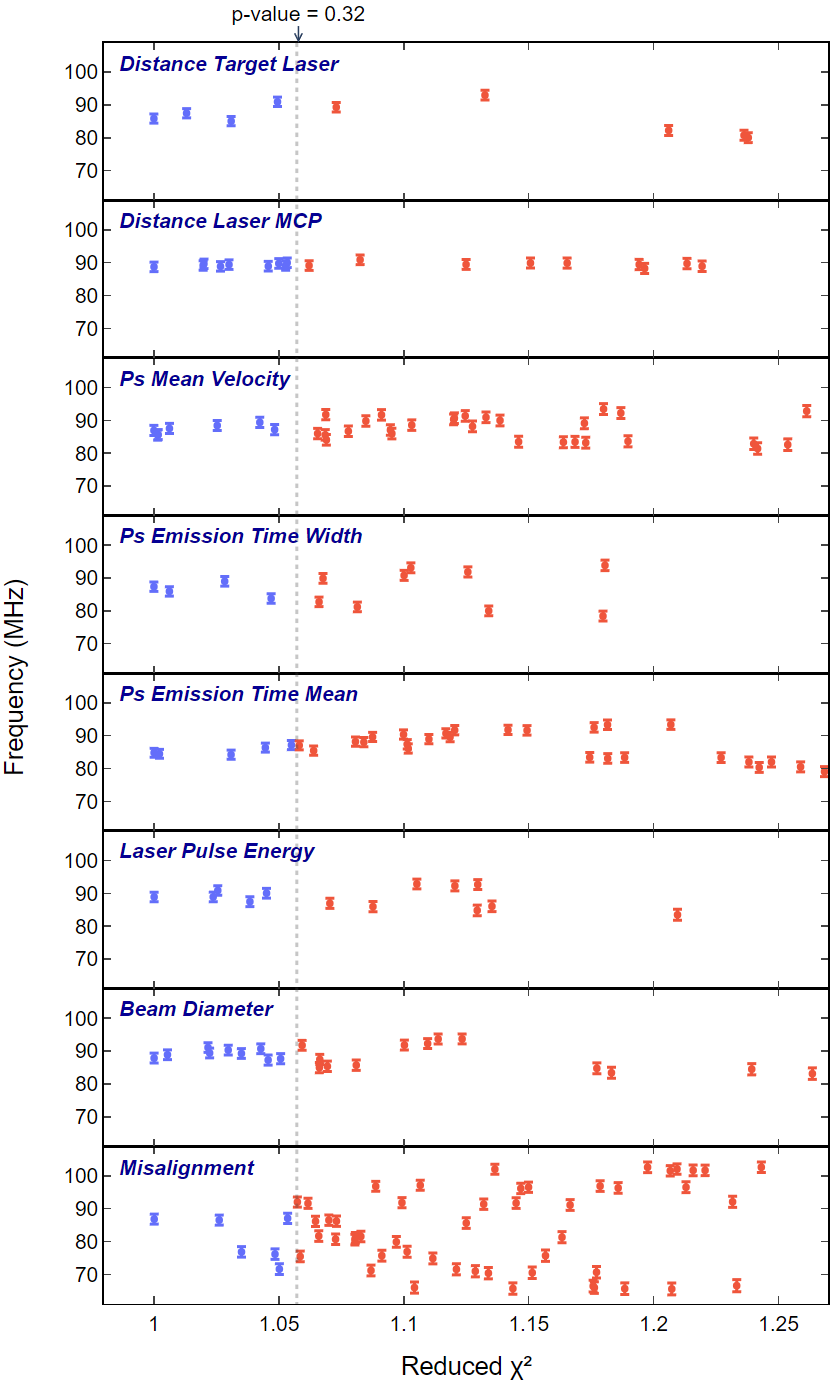} 
\caption[Plot of reduced $\chi^2$ values for varying individual simulation parameters of run 1.]{Plot of reduced $\chi^2$ values for varying individual simulation parameters of run 1. The cut-offs for a p-value of 0.32 are shown as dotted lines. Fits below correspond to be within a $68\%$ confidence interval and are taken into account for the estimation of the magnitude of the parameters influence on the systematic error.
\label{fig:relative-sys}}
\end{figure}

Figure \ref{fig:relative-sys} shows the respective reduced $\chi^2$ values for scans of the individual parameters. The extracted systematic errors can be found in table \ref{tab:errors_sys}. Note that this method is limited in scope, since individual errors are in general not uncorrelated. A full scan of the parameter space (see figure \ref{fig:chi2-minimization}) is necessary to extract a reliable estimate on the overall error. Furthermore, the estimation includes statistical errors due to the fitting procedure of the simulation output and will therefore be overestimated, especially for smaller contributions. However, these results allow to compare the relative impact of the parameter uncertainties on the systematic error.

\begin{table} [ht!]
  \begin{tabular}
    {| c || c |}
    \hline
    Magnitude & Parameter \\
    \hline
    \hline
    $7.7\,\mathrm{MHz}$ & Reflection Misalignment \\
    $2.9\,\mathrm{MHz}$ & Laser Position (relative to target) \\
    $2.6\,\mathrm{MHz}$ & Positronium Emission Time Width \\
    $1.9\,\mathrm{MHz}$ & Positronium Mean Velocity \\
    $1.9\,\mathrm{MHz}$ & Laser Beam Waist Size \\
    $1.7\,\mathrm{MHz}$ & Laser Pulse Energy \\
    $1.5\,\mathrm{MHz}$ & Positronium Emission Time Mean \\
    $0.6\,\mathrm{MHz}$ & MCP Position (relative to laser) \\
    \hline
   \end{tabular} 
\caption{Sources of systematic fitting error for run 1.
\label{tab:errors_sys}}
\end{table}

The most significant contribution to the systematic error stems from the residual first order Doppler shift, which is directly proportional to the misalignment of the incoming and retro-reflected beam. Since it also scales linearly with positronium velocity, it is also partly responsible for the contribution from the timing and velocity profile of positronium emission.

Due to the short pulse length (relative to the transit time), the excitation laser preferentially probes a subset of velocities given by its spatial and temporal properties. This results in the second order Doppler shift contributing to the systematic error present on those parameters, as well as on parameters related to the emission profile. To a lesser degree, the AC-Stark shift similarly leads to systematic errors in parameters related to the spatial and temporal intensity profile seen by the positronium atoms, mainly due to uncertainties on the precise beam profile and pulse energy.

\begin{figure}
\includegraphics[width=\columnwidth]{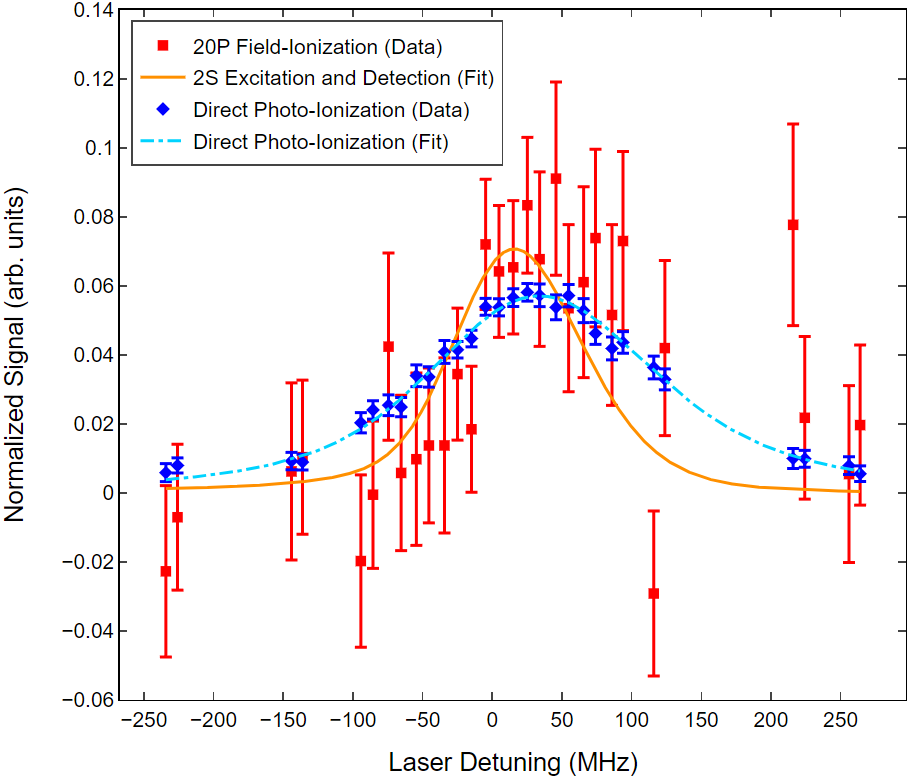} 
   \caption[Positronium direct photo-ionization and 2S excitation measured via field-ionization as a function of laser detuning for data taken during run 2.]{Positronium direct photo-ionization (blue diamonds) and 2S excitation measured via field-ionization (red squares) as a function of laser detuning for data taken during run 2. The data is sum-normalized and the best fits for all simulated parameter sets for run 2 is shown (light-blue dashdotted line for direct photo-ionization and orange solid line for 2S excitation).   
\label{fig:analysis-line-fit-run2}}
\end{figure}

\begin{figure}
\includegraphics[width=\columnwidth]{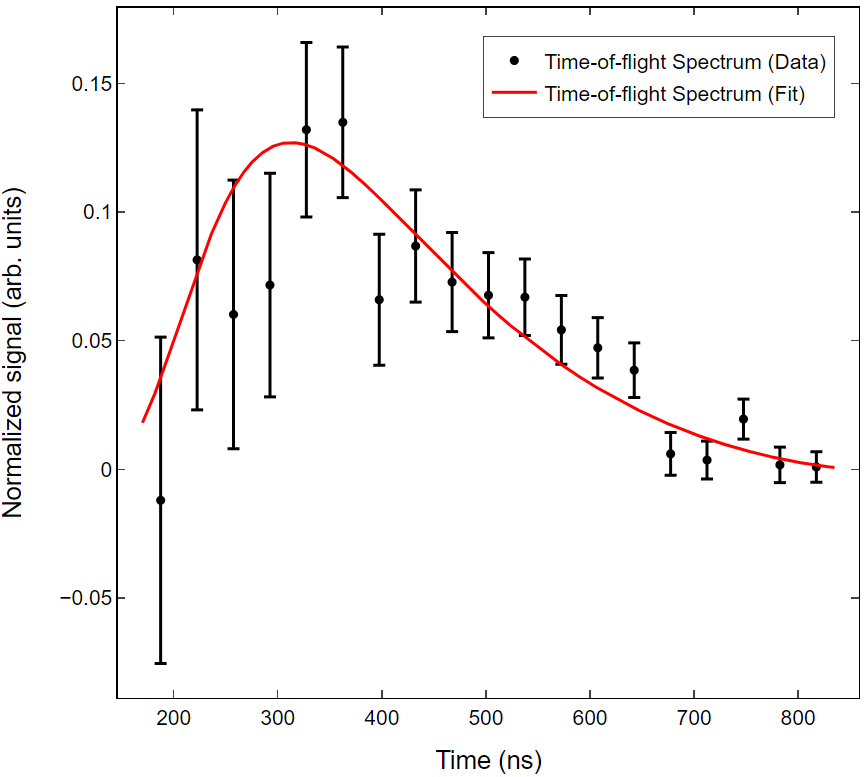} 
   \caption[Normalized time-of-flight spectrum and best fit for all simulated parameter sets for run 2.]{Normalized time-of-flight spectrum relative to the timing of the 20P excitation to field-ionization of the Rydberg atoms on the MCP and the corresponding best fit for all simulated parameter sets for run 2.
\label{fig:analysis-tof-fit-run2}}   
\end{figure}

\paragraph{Results of run 2:} Figure \ref{fig:analysis-line-fit-run2} and \ref{fig:analysis-tof-fit-run2} shows the dataset gathered during run 2 after chirp correction and re-binning, in addition to the best fit determined by $\chi^2$ minimization analogous to run 1. Unfortunately, the statistical error is significantly larger in this dataset. Firstly, the supplied positron rate is approximately a factor 2 smaller than for run 1, due to the decay of the sodium source. Furthermore, the solid angle acceptance of the MCP for field-ionization is approximately a factor 2 smaller, in addition to higher losses of the grids due to the shallower angle of incidence. Lastly, we found that the quantum efficiency of the detector for photons from positron decay was significantly higher in this arrangement, most likely due to the increased path length at an angle of $45^\circ$. This further decreased the signal-to-noise ratio for this dataset.

The estimation of the systematic error proceeds analogously to run 1. However, due to larger errors in the data points, the line-shape and time-of-flight spectrum cannot be sampled as precisely and do not have the same impact on the quality-of-fit parameter. Therefore, the error estimate is correspondingly higher than in the case of run 1. The final result for the resonance frequency in terms of laser detuning obtained reads
\begin{equation}
 616\,803\,599.2\:\mathrm{MHz} \pm 26.9\:\mathrm{MHz}
\end{equation}
where the respective sources of error can be found in table \ref{tab:errors_run2}.

\begin{figure}
\includegraphics[width=\columnwidth]{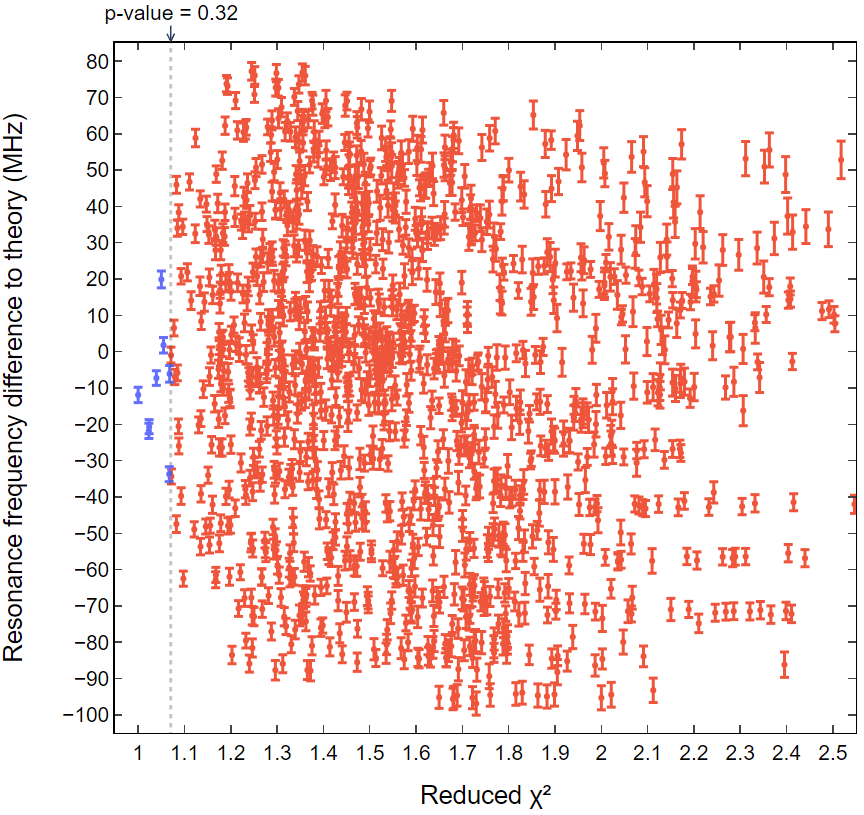} 
   \caption[Plot of reduced Chi-Squared$\chi^2$ values for the fits of the dataset of run 2 to simulated line-shapes and TOF spectra.]{Plot of reduced $\chi^2$ values for the fits of the dataset of run 2 to simulated line-shapes and TOF spectra. The cut-off for a p-value of 0.32 is shown as a dotted line. Fits below correspond to be within a $68\%$ confidence interval and are taken into account for the estimation of the systematic error.
\label{fig:chi2-minimization-run2}}
\end{figure}
 
\begin{table} [ht!]
  \begin{tabular}
    {| c || c |}
    \hline
    Magnitude & Description \\
    \hline
    \hline
    $2.1\,\mathrm{MHz}$ & Data Fitting Error  \\
    $0.8\,\mathrm{MHz}$ & Simulation Fitting Error  \\
    $26.8\,\mathrm{MHz}$ & Systematic Fitting Error  \\
    $1.0\,\mathrm{MHz}$ & Chirp Correction Error \\
    $0.2\,\mathrm{MHz}$ & Absolute Frequency Reference Error \\
    \hline
    $26.9\,\mathrm{MHz}$ & Quadrature Sum \\
    \hline
   \end{tabular} 
\caption{Sources and magnitude of statistical and systematic errors present in result for run 2.
\label{tab:errors_run2}}
\end{table}

This corresponds to a transition frequency of
\begin{equation}
 1\,233\,607\,198.4\:\mathrm{MHz} \pm 53.8\:\mathrm{MHz}
\end{equation}
or equivalently, this result differs from theory by
\begin{equation}
 -23.8\,\mathrm{MHz} \pm 53.8\,\mathrm{MHz} \pm 0.6\,\mathrm{MHz} \quad \text{.}
\end{equation}

In summary, the addition of a frequency comb (see section \ref{sec:frequency-stabilization}), as well as a pulse-by-pulse determination of frequency chirp allowed to significantly reduce the main systematic errors present in run 1. This resulted in increased precision, even in a situation with a significant reduction in signal-to-noise ratio.

\paragraph{Combined result and summary:}

Assuming uncorrelated measurements, both results can be combined using the maximum likelihood estimator for the mean \cite{Bevington2003} given by
\begin{equation}
 \bar \mu = \frac{\sum \frac{x_i}{{\sigma_i}^2}}{\sum \frac{1}{{\sigma_i}^2}} 
\end{equation}
where the uncertainty of the combined value is given by
\begin{equation}
 \sigma_{\bar \mu} = \sqrt{\frac{1}{\sum \frac{1}{{\sigma_i}^2}}} \quad \text{.}
\end{equation}

The weighted average for the resonance frequency in terms of laser detuning then yields
\begin{equation}
 616\,803\,605.2\:\mathrm{MHz} \pm 24.8\:\mathrm{MHz}
\end{equation}
which corresponds to a transition frequency of
\begin{equation}
 1\,233\,607\,210.5\:\mathrm{MHz} \pm 49.6\:\mathrm{MHz}
\end{equation}
or equivalently, this result differs from theory by
\begin{equation}
 -11.7\,\mathrm{MHz} \pm 49.6\,\mathrm{MHz} \pm 0.6\,\mathrm{MHz} \quad \text{.}
\end{equation}

\begin{figure}
\includegraphics[width=\columnwidth]{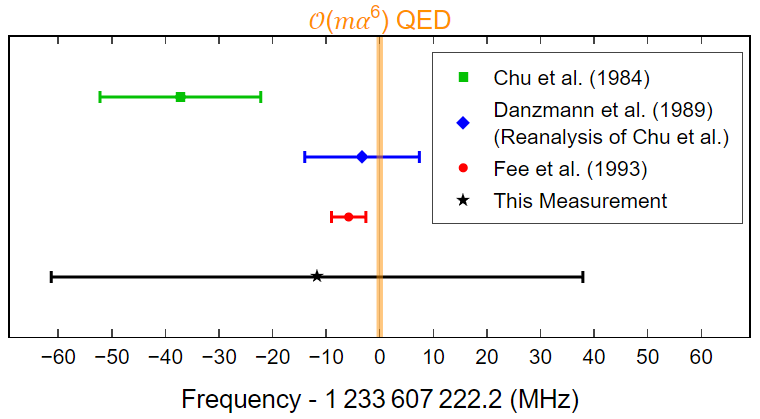} 
   \caption[Summary of 1S-2S experimental results and most recent QED theory value and uncertainties.]{Summary of 1S-2S experimental results and most recent QED theory value and uncertainties. The dataset of Chu et al. (green square) \cite{Chu1984} was re-analyzed by Danzmann et al. (blue diamond) \cite{Danzmann1989}. The currently most precise experimental value was measured by Fee et al. \cite{Fee1993} (red circle). The shaded orange/grey band corresponds to the theoretical QED prediction at $\mathcal{O}\left( m\alpha^6 \right)$ \cite{Pachucki1998} including an error of $0.6\,\mathrm{MHz}$.
\label{fig:result-1s2s}}
\end{figure}

Figure \ref{fig:result-1s2s} shows the result presented here in the context of the most recent QED theory value and previous experimental values.

\section{Conclusions}

As outlined in the introduction, positronium represents an ideal system to probe for new physics beyond the Standard Model. Spectroscopic precision measurements allows one to stringently test bound-state Quantum Electrodynamics, without the complications presented by the finite size effect of protonic atoms. Especially the advent of pulsed slow positron beams has led to many advancements in this field \cite{RevModPhys.87.247}. 

Over the last few decades, measurements in positronium have been a driver for many theoretical developments and vice versa. However, the currently most precise measurement, which is the determination of the $1^3\text{S}_1 \to 2^3\text{S}_1$ transition by Fee et al. \cite{Fee1992} at a level of $2.6\,\mathrm{ppb}$, is still almost an order of magnitude less precise than the most recent calculations. Our results at the $40\,\mathrm{ppb}$ are not yet competitive with the best determination. However, our technique combined with CW laser excitation \cite{Cooke2015} will result in an improvement of at least an order of magnitude compared to the current state of the art (see details in Appendix \ref{sec:appendix-ProspectCW}). This will allow to check QED calculations up to the current order $m\alpha^6$ and motivate further developments towards a full calculation of order $m\alpha^7$. Moreover, the recent demonstration of positronium laser cooling \cite{AEgIS:2023lpw, Shu:2023fgm}, gives great prospects for a new generation of high precision spectroscopy with this unique atomic system.

\section*{Acknowledgements}
We are in debt with A. Antognini, K. Kirch and F. Merkt for their continuous and essential support on this project. We acknowledge D. Cooke and G. Wichmann for their substantial contributions to the development of the apparatus which provided the basis for these results. We are grateful to A. Soter and D. Yost and B. Radics for their very useful comments and discussions. 
This work was supported by the ERC consolidator grant 818053-Mu-MASS  and the Swiss National Science Foundation under grant 166286. 

\appendix

\section{\label{sec:appendix-tau}Rydberg state selection}

The lifetime of excited states with $n>1$ is limited both by annihilation and radiative decay, such that
\begin{equation}
 \tau_{\text{n}^{(2\text{S}+1)}\text{L}_\text{J}} = \left[ \Gamma^a_{\text{n}^{(2\text{S}+1)}\text{L}_\text{J}} + \Gamma^r_{\text{n}^{(2\text{S}+1)}\text{L}_\text{J}} \right]^{-1}
 \label{eq:state-lifetime}
\end{equation}
where $\Gamma^a$ is the decay rate due to annihilation and $\Gamma^r$ is the radiative decay rate. The annihilation cross-section crucially depends on the overlap of the wave-functions of the positron and the electron. Therefore, the lifetime of S states ($\text{L}=0$) scales with $\text{n}^3$ \cite{Ore1949}. For the same reason annihilations are highly suppressed for states with $\text{L}>0$. The shortest annihilation lifetime of such a state \cite{Sen2019} is  
\begin{equation}
 \tau^a_{2^3\text{P}_2} = 99.57\times10^{-6}\,\mathrm{s}
 \label{eq:2p-lifetime}
\end{equation}
which is still two orders of magnitude higher than
\begin{equation}
 \tau^a_{2^3\text{S}_1} = 2^3 \cdot 142.05\times10^{-9}\,\mathrm{s} = 1.136 \times10^{-6}\,\mathrm{s}
\end{equation}
for the equivalent S state. Annihilation lifetimes of P states scale with ${n^5}/{(n^2-1)}$ \cite{Sen2019}, which for large n becomes approximately $n^3$, analogously to the S states.

The radiative decay rate is given by the sum of the Einstein coefficients of spontaneous decay to all lower lying states \cite{Gallagher1994}, such that 
\begin{equation}
\begin{aligned}
 \Gamma^r_{\text{n}^{(2\text{S}+1)}\text{L}_\text{J}} &= \sum A_{\text{n'L'}\to\text{nL}} \\ &=\sum \frac{16 \pi \alpha \left(E_{\text{n'L'}\to\text{nL}}\right)^3}{\hbar^3 c^2} \\ &\quad \times \left| \braket{\text{n'L'}|r|\text{nL}} \right|^2
 \label{eq:einstein-ps}
\end{aligned}
 \end{equation}
where $E_{\text{n'L'}\to\text{nL}}$ is the transition energy and $\braket{\text{n'L'}|r|\text{nL}}$ the transition matrix element.

\begin{figure}
\includegraphics[width=\columnwidth]{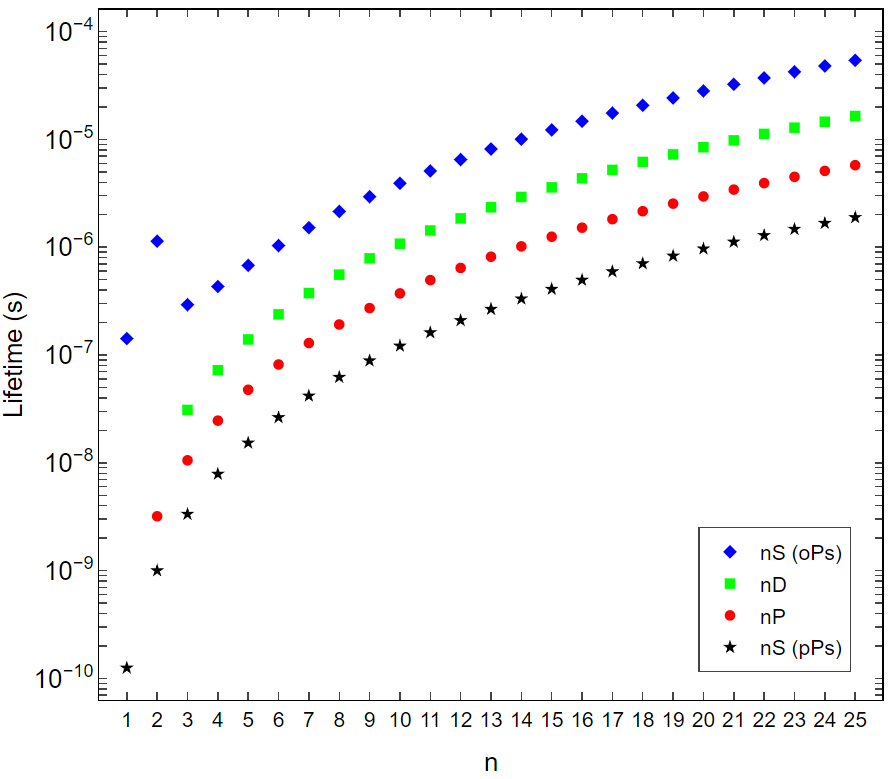}
\caption[Lifetime for positronium states with $\text{n}<26$ including annihilation and radiative decay.]{Lifetime for positronium states with $\text{n}<26$ including annihilation (for $\text{n}<3$) and radiative decay. Numerical values for radiative lifetimes were taken from \cite{Jitrik2004} and scaled by a factor 2 to account for the reduced mass.\label{fig:positronium-lifetime}}
\end{figure} 

Figure \ref{fig:positronium-lifetime} shows the lifetime for S ($\text{L}=0$), P ($\text{L}=1$) and D ($\text{L}=2$) states. For S states with $\text{n}>1$, the lifetime of pPs is dominated by annihilation due to it decaying predominantly to two photons, while oPs is mainly governed by radiative decay. However, an important exception is oPs in the 2S state, which is radiatively meta-stable. In fact, due to the small fine-structure splitting, the decay rate to 2P is minuscule and the main contribution is the electric dipole transition directly to the 1S, by the simultaneous emission of two photons \cite{GoeppertMayer1931}. For P and D states the total spin does not play a significant role, since the radiative decay rates are many orders of magnitude larger than the respective annihilation decay rates. 

A competing loss mechanism for detecting Rydberg states are collisions on background gas atoms. Assuming the positronium atom is lost upon collision with a background gas atom, the probability for it to reach the MCP is
\begin{equation}
 P_{\text{col}}(l) = e^{-\frac{l}{\lambda}}
\end{equation}
where $l$ is the length travelled by positronium and $\lambda$ is the mean free path \footnote{The usual relative velocity correction of $\sqrt{2}$ in the denominator of the mean free path is not included, since background gas atoms appear essentially stationary for positronium.}
\begin{equation}
 \lambda = \frac{k_B T}{p \, \sigma_{\text{col}}}
\end{equation}
with $k_B$ being the Boltzmann constant, $T\approx300\,\mathrm{K}$ is the temperature and $p\approx10^{-8}\,\mathrm{mbar}$ the pressure of the background gas and $\sigma_{\text{col}}$ the collision cross-section.

The cross-section $\sigma_{\text{col}}$ can be estimated by assuming elastic collisions of two particle species with different radii, such that
\begin{equation}
 \sigma_{\text{col}} = \pi \left(r_{gas} + r_{Ps}\right)^2
\end{equation}
where $r_{Ps}$ is the radius of positronium and $r_{gas}$ is the average radius of background gas molecules. The main gas species in the chamber is nitrogen reaching the chamber from the trap, for which the kinetic radius is $r_{\text{N}_2} = 1.8\times10^{-10}\,\mathrm{m}$ \cite{Ismail2015}. For positronium the RMS radius
\begin{equation}
 r_{\text{RMS}} = 2 a_0 \, \text{n} \, \sqrt{\frac{5\text{n}^2-3\text{L}(\text{L}+1)+1}{2}} \quad \text{,}
\end{equation}
which corresponds to the RMS radius of hydrogen \cite{Shankar1994} scaled with the reduced mass of positronium, is an appropriate approximation. We therefore find
\begin{equation}
 \sigma^{\text{nP}}_{\text{col}}(\text{n}) \approx \left(3.19 + 2.96 \,\text{n}\,\sqrt{\text{n}^2-1}\right)^2 \times 10^{-20}\,\mathrm{m}^2
\end{equation}
for P states with arbitrary n.

\begin{figure}
\includegraphics[width=\columnwidth]{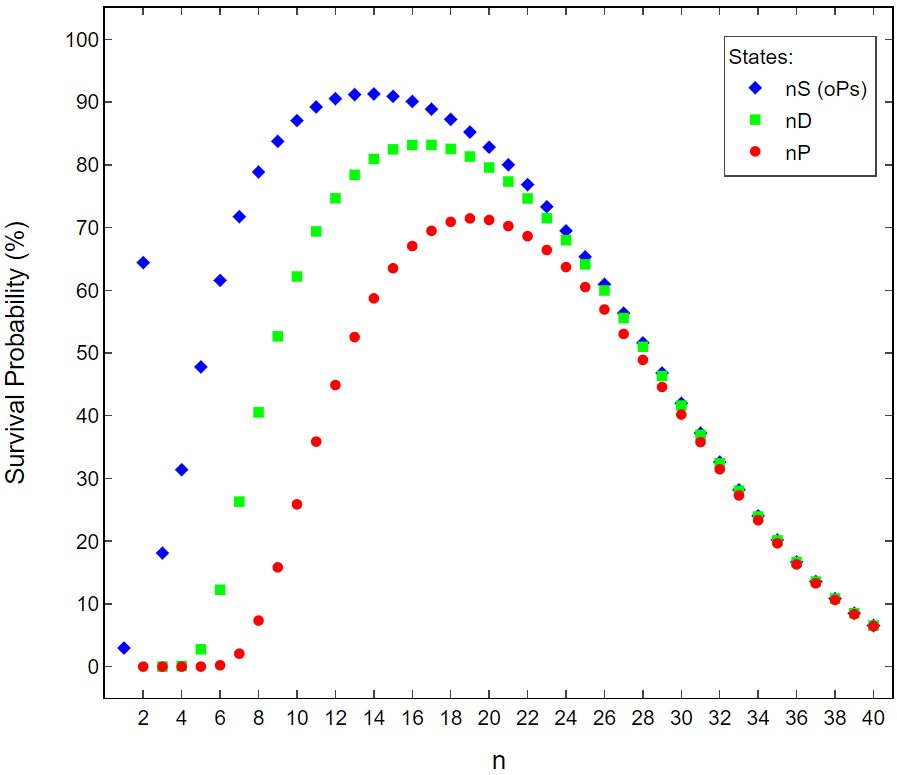} 
   \caption[Survival probability of positronium for different principal quantum numbers in S, P and D states.]{Survival probability of positronium for different principal quantum numbers in S, P and D states travelling $5\,\mathrm{cm}$ at a velocity of $10^5\,\frac{\mathrm{m}}{\mathrm{s}}$ in vacuum with a background gas pressure of $10^{-8}\,\mathrm{mbar}$.
   \label{fig:rydberg-survival}}        
\end{figure}

Combining the losses due to collisions with the decay probability yields
\begin{equation}
 P^{\text{nP}}(l,\text{n}) = P^{\text{nP}}_{\text{col}}(l,\text{n}) \cdot e^{-\frac{l}{v_{\text{Ps}}\,\tau_{\text{nP}}}}
\end{equation}
for the probability of an atom with velocity $v$ and lifetime $\tau_{\text{nP}}$ to reach the detector in the same state, after flying a distance of $l$. Figure \ref{fig:rydberg-survival} shows the survival probability for a pressure of $p=10^{-8}\,\mathrm{mbar}$, velocity of $v_{\text{Ps}}=10^5\,\frac{\mathrm{m}}{\mathrm{s}}$ and distance $l=5\,\mathrm{cm}$, which corresponds to the approximate experimental parameters of both runs. The highest rate is therefore expected for n in the range from 18-20. 

The final consideration for which specific n to use is the efficiency of field ionization on the detector. The field strengths achieved on the detector assembly are on the order of 2-4 ${\mathrm{kV}}/{\mathrm{cm}}$. For these field strengths, the probability of ionization has been shown to be close to unity for $\text{n}>19$ \cite{Alonso2018}. Therefore, the 20P state was chosen for the Rydberg detection scheme.

\section{\label{sec:appendix-MC}Monte-Carlo simulation and fitting procedure}

\subsection{Optical Bloch equations\label{sec:lasercalc}}

To calculate the probability of exciting ground-state positronium to the 2S state, a density matrix approach for a two-level system \cite{Haas2006} is used. This is done by solving the set of differential Bloch equations

\begin{equation}
\begin{aligned}
	\partial_t \, \rho_{gg}(t) &= - \Omega \, \operatorname{Im}(\rho_{ge}(t)) - \Gamma_{1S} \, \rho_{gg}(t) \\
	\partial_t \, \rho_{ge}(t) &= -i \, \Delta \omega \, \rho_{ge}(t) + \frac{i \, \Omega}{2}  (\rho_{gg}(t) - \rho_{ee}(t)) \\
     & - \frac{\Gamma_i + \Gamma_{1S} + \Gamma_{2S}}{2} \, \rho_{ge}(t) \quad \text{,} \\
	\partial_t \, \rho_{ee}(t) &= \Omega \, \operatorname{Im}(\rho_{ge}(t)) - (\Gamma_i + \Gamma_{2S}) \, \rho_{ee}(t) 
\end{aligned}
\label{eq:density-diffeq}
\end{equation}
assuming that all positronium atoms are initially in the ground state. 
Equations (\ref{eq:density-diffeq}) can be integrated numerically by using a suitable algorithm. To arrive at the simulation results presented in this article (see section \ref{subsec:simlaser}) an adaptive Runge-Kutta method \cite{Press2005} was employed. 

$\rho_{gg}$ represents the state population of the ground state, while $\rho_{ee}$ is the state population of the excited state. The annihilation decay rates read
\begin{equation}
\Gamma_{1S} = \frac{1}{\tau^a_{1^3\text{S}_1}} \approx \frac{1}{142\, \mathrm{ns}}
\label{eq:decayrate1s}
\end{equation}
and
\begin{equation}
\Gamma_{2s} = \frac{1}{\tau^a_{2^3\text{S}_1}} \approx \frac{1}{1.14\, \mathrm{\mu s}}
\label{eq:decayrates}
\text{ .}
\end{equation}

The spontaneous decay of the 2S states back into the ground state does not need to be included, since the the annihilation rate $\Gamma_{2s}$ is many orders of magnitude larger. Ionization, however, needs to be included in the model, especially for higher power lasers, since the rate 
\begin{equation}
\Gamma_i = 2\pi \, \beta_i \, I(t)
\label{eq:ionirate}
\end{equation}
is directly proportional to the laser intensity $I(t)$. The ionization coefficient $\beta_i$ is given by its value for hydrogen, taken from Haas et al. \cite{Haas2006}, after multiplying the proper rescaling factor
\begin{equation}
\begin{aligned}
  \beta_i & = \frac{1}{Z^4} \left(\frac{m_e}{\mu}\right)^3 \cdot 1.20208 \times 10^{-4} \,\frac{\mathrm{Hz}\,\mathrm{m}^2}{\mathrm{W}} \nonumber \\ 
&= 9.61664 \times 10^{-4} \,\frac{\mathrm{Hz}\,\mathrm{m}^2}{\mathrm{W}}
\text{ ,}
\label{eq:rescale-ioni}  
\end{aligned}
\end{equation}
where $Z=1$ is the nuclear charge number and $\mu = \frac{m_e}{2}$ is the reduced mass.
 
The two-photon Rabi frequency 
\begin{equation}
\Omega = 2\left(2\pi\,\beta_{ge}\right) I(t)
\label{eq:rabifreq}
\end{equation}
used in equation (\ref{eq:density-diffeq}) is proportional to the laser intensity $I(t)$ and a rescaled two-photon transition matrix element \cite{Haas2006}
\begin{equation}
\begin{aligned}
\beta_{ge} &= \frac{1}{Z^4} \left(\frac{m_e}{\mu}\right)^3 \cdot 3.68111 \times 10^{-5} \,\frac{\mathrm{Hz}\,\mathrm{m}^2}{\mathrm{W}} \nonumber \\
&= 2.944888 \times 10^{-4} \,\frac{\mathrm{Hz}\,\mathrm{m}^2}{\mathrm{W}}
\text{ .}
\label{eq:rescale-transmatrix}
\end{aligned}
\end{equation}

Finally, the excitation detuning is given by
\begin{equation}
\begin{aligned}
	\Delta \nu &= 2 \, \nu_L - \nu_0 \left(1 + \frac{\vec v \cdot \left(\vec{e}_i + \vec{e}_r \right)}{2c} - \frac{v^2}{2c^2} \right)  \\
    &- \Delta\nu^e_{ac} + \Delta\nu^g_{ac}
\end{aligned}
\label{eq:detuning}
\end{equation}

where $\nu_L$ is the frequency of the laser used for excitation, $\nu_0$ is the unperturbed transition frequency, which is shifted by the relativistic Doppler effect (shown to second order) and the AC Stark shift. The Doppler shift is given by the the positronium velocity $\vec{v}$ and the propagation axes of the incoming ($\vec{e}_i$) and reflected ($\vec{e}_r$) photons. In the case of perfectly overlapping retro-reflected laser beam profiles, $\vec{e}_i = - \vec{e}_r$ and the first order term cancels. Experimentally, this can be realized by employing a resonant Fabry-Perot cavity. Otherwise, there will be some residual first-order Doppler shift, depending on the angle between the counter-propagating wave fronts on the order of
\begin{equation}
\Delta\nu_{d1} \approx \pm 200 \, \frac{\mathrm{kHz}}{\mathrm{\mu rad}}
\label{eq:firstdoppler}
\text{ .}
\end{equation}

The AC Stark shifts are given by
\begin{equation}
\Delta\nu^{e/g}_{ac} = \beta^{e/g}_{ac} \, I(t)
\text{ .}
\label{eq:starkshift}
\end{equation}
Analogous to the ionization coefficient and the transition matrix element, the AC Stark shift coefficients $\beta^{e/g}_{ac}$ are taken from Haas et al. \cite{Haas2006}, after multiplying the rescaling factor to yield
\begin{equation}
\begin{aligned}
\beta^e_{ac} &= \frac{1}{Z^4} \left(\frac{m_e}{\mu}\right)^3 \cdot 1.39927 \times 10^{-4} \,\frac{\mathrm{Hz}\,\mathrm{m}^2}{\mathrm{W}} \\
& = 1.119416 \times 10^{-3} \,\frac{\mathrm{Hz}\,\mathrm{m}^2}{\mathrm{W}}
\end{aligned}
\label{eq:rescale-starke}
\end{equation}
and
\begin{equation}
\begin{aligned}
\beta^g_{ac} & = \frac{1}{Z^4} \left(\frac{m_e}{\mu}\right)^3 \cdot 2.67827 \times 10^{-5} \,\frac{\mathrm{Hz}\,\mathrm{m}^2}{\mathrm{W}} \\
&= 2.142616 \times 10^{-4} \,\frac{\mathrm{Hz}\,\mathrm{m}^2}{\mathrm{W}}
\text{ .}
\end{aligned}
\label{eq:rescale-starkg}
\end{equation}

Due to the relatively high velocities of positronium atoms produced with porous silica targets, they interact with the laser only for a very short period of time and the spatial and temporal profiles of the laser beam need to be taken into account. Assuming sufficiently collimated beams with Gaussian spatial profiles, the intensity can be written as
\begin{equation}
I(y,z) = \frac{2 \, P}{\pi \, w} \, e^{-2 \, \frac{y^2+z^2}{{w}^2}}
\text{ ,}
\label{eq:intensity}
\end{equation}
where $y$ and $z$ are the time dependent coordinates of the positronium atom, with the propagation axis of the laser assumed to be parallel to the $x$-axis. The waist size $w$ corresponds to the $1/e^2$-radius of the laser beam. While the Power $P$ is assumed to be constant for a CW laser, for a pulsed laser a non-trivial temporal profile needs to be taken into account. It is given by a combination of the intrinsic Gaussian time profile of the incoming laser pulse
\begin{equation}
P_i(t) = \frac{E_p}{\sqrt{2\pi} \,  \sigma} \, e^{-\, \frac{\left(t-t_{\mathrm{off}}\right)^2}{2\,  \sigma^2}}
\label{eq:pulsedpowerin}
\end{equation}
and the reflected laser pulse
\begin{equation}
P_r(t) = \frac{E_p}{\sqrt{2\pi} \,  \sigma} \, e^{-\, \frac{\left(t+t_{\mathrm{ref}}-t_{\mathrm{off}}\right)^2}{2\,  \sigma^2}}
\text{ ,}
\label{eq:pulsedpowerref}
\end{equation}
where $E_p$ is the pulse energy, $t_{\mathrm{off}}$ is the timing offset between positronium emission and the peak intensity of the laser pulse and $t_{\mathrm{ref}}$ is the round-trip time of the laser pulse to the mirror and back.

The combination required is the trivial superposition
\begin{equation}
P(t) = P_i(t) + P_r(t)
\label{eq:pulsedpowerone}
\end{equation}
when used for the calculation of the ionization and Stark shift coefficients, since these are one-photon processes. However, excitation is strictly a two-photon resonant effect. This is taken into account by using the envelope of the two counter-propagating beams, which results in the expression
\begin{equation}
P(t) = 2 \sqrt{P_i(t) P_r(t)}
\text{ .}
\label{eq:pulsedpowertwo}
\end{equation}

\subsection{Description of Monte Carlo simulation\label{sec:lasersim-desc}}

A Monte Carlo simulation was developed to accurately reproduce the lineshape for a given set of experimental parameters. Due to the slightly different experimental setup and improved measurement system run 1 and 2 use a different set of input parameters, which can be found in tables \ref{tab:lasersimvars_run1} and \ref{tab:lasersimvars_run2}. 

\begin{figure}
\includegraphics[width=\columnwidth]{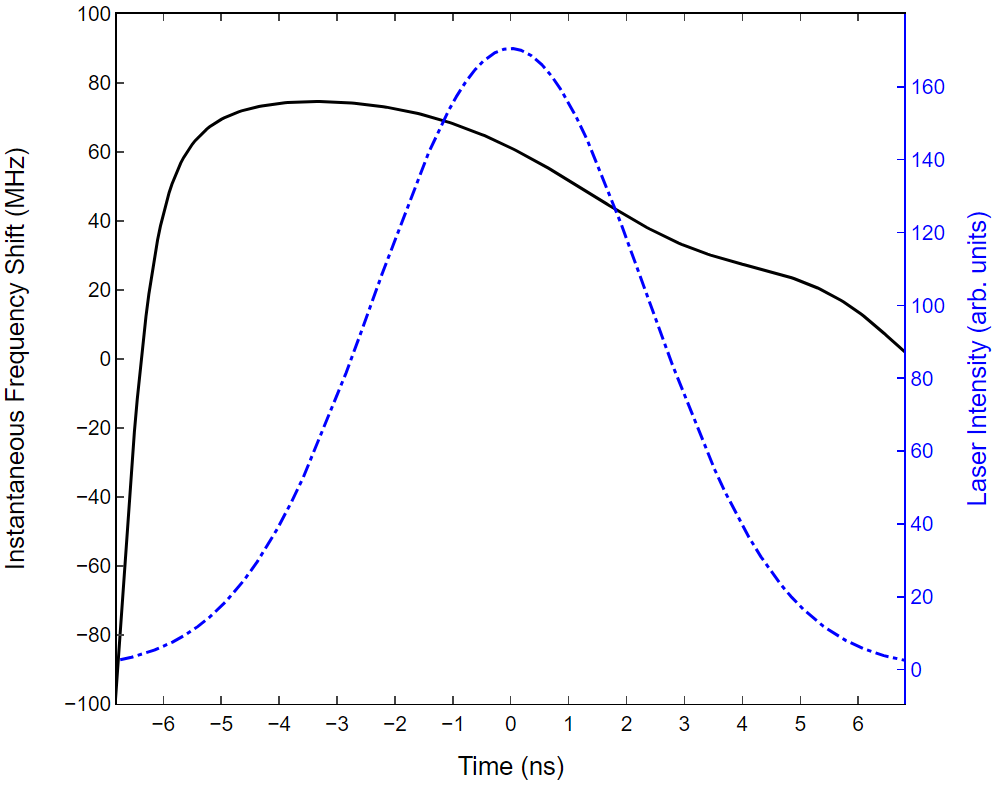} 
   \caption[Average chirp profile and simulated laser intensity time profile]{Average chirp profile (solid black) as taken during run 2 and simulated laser intensity time profile (dashdotted blue) for a time spread of $5.5\,\mathrm{ns}$ (FWHM).   
\label{fig:chirp-sim}}
\end{figure}

The effects of the laser chirp are included by modifying the instantaneous frequency according to the average of the chirp measurements taken during run 2. Figure \ref{fig:chirp-sim} shows the average shift from the laser detuning during the pulse length and the mean Gaussian time profile of the simulated intensity profile. This chirp profile is corrected by the intensity-weighted average chirp and amounts to
\begin{equation}
 \overline{\Delta \nu_c} = 56.6\,\mathrm{MHz}
\end{equation}
for the data shown in figure \ref{fig:chirp-sim}.

\begin{table*} [ht!]
  \begin{tabular}
    {| c || c || c |}
    \hline
    Variable Name & Typical Value & Description \\
    \hline
    \hline
    WaistSize & 2 mm & laser beam minimum waist radius \\ 
    PulseEnergy & 6.25 mJ & laser pulse energy \\
    PulseFWHM & 6.7 ns & FWHM of temporal pulse shape \\
    LaserTimeOffset & 28 ns & mean difference of Ps emission to laser time \\
    MisalignmentY & 0 mrad & angle between laser wave-fronts (XY-plane) \\
    MisalignmentZ & 2 mrad & angle between laser wave-fronts (XZ-plane) \\
    PositroniumTemp & 700 K & temperature for velocity distribution \\
    DistanceSourceLaserX & 150 mm & X-axis distance target / mirror \\
    DistanceSourceLaserY & 0 mm & Y-axis distance target / laser \\
    DistanceSourceLaserZ & 3 mm & Z-axis distance target / laser \\
    SigmaBeamX & 1.5 mm & X-axis spread of Ps emission area \\
    SigmaBeamY & 1.5 mm & Y-axis spread of Ps emission area \\
    SigmaEmission & 20 ns & Ps emission time width \\
    DistanceMCPLaserX & 0 mm & X-axis distance laser / MCP \\
    DistanceMCPLaserY & 0 mm & Y-axis distance laser / MCP \\
    DistanceMCPLaserZ & 38 mm & Z-axis distance laser / MCP \\
    TargetAngle & $0{}^\circ$ & target angle relative to positron beam \\
    \hline
   \end{tabular} 
\caption{Experimental variables and typical values used in Monte-Carlo simulation (run 1)
\label{tab:lasersimvars_run1}}
\end{table*}

\begin{figure*}
\includegraphics[width=0.92\textwidth]{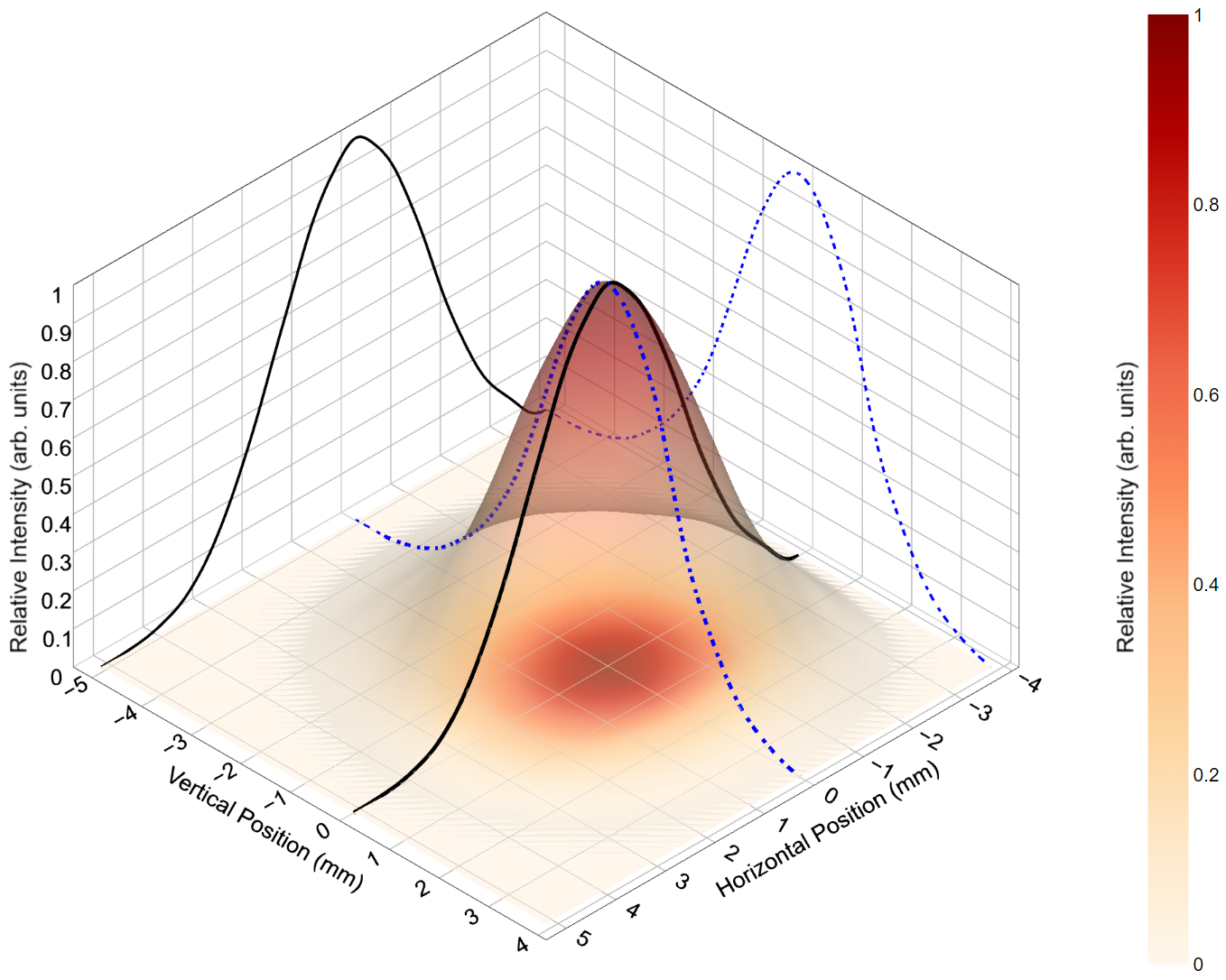} 
   \caption[Spatial laser beam profile taken during run 2]{3D surface plot of spatial laser beam profile taken during run 2, with 2D color space projection on the z-plane and 2D contour projections of the beam center for the horizontal (solid black) and vertical (dashdotted blue) planes.
\label{fig:spatial_laser_beam_profile}}
\end{figure*}

Additionally, for run 2 the simulation is supplied with the spatial laser beam profile taken with a Thorlabs BP209-VIS beam profiler (see figure \ref{fig:spatial_laser_beam_profile}). It was found to be a slightly skewed elliptical Gaussian beam with waist sizes
\begin{align}
 w_\mathrm{hor} = (2.91 \pm 0.01) \, \mathrm{mm} \\
 w_\mathrm{ver} = (2.58 \pm 0.01) \, \mathrm{mm} 
\end{align}
where the vertical axis lies parallel and the horizontal axis perpendicular to the target surface. However, due to pointing instability and the averaging nature of the measurement with a rotating slit beam profiler the true beam diameter is probably slightly lower. This has been taken into account in the simulation by including correction factors for the vertical and horizontal beam size (see table \ref{tab:lasersimvars_run2}).

For each simulated atom, the primary positronium vertex is created according to an elliptical Gaussian profile of widths SigmaBeamX and SigmaBeamY. The coordinate system is chosen such that the X-axis corresponds to the propagation axis of the laser, while the Z-axis lies perpendicular to the target. The implantation and emission time profile is approximated by varying the creation time of the atom using a $\mathcal{N}(0,\,\text{SigmaEmission}^2)$ distribution.

The initial velocity vector is created with a $\cos \theta$-distribution \cite{Greenwood2002} for the direction and a Maxwell-Boltzmann distribution \cite{Mohamed2011} for the magnitude. It therefore reads
\begin{align}
 {v^x}_i &= |v|_i \cdot \sqrt{{ST}_i} \cdot \cos(2\pi P_i) \\
 {v^y}_i &= |v|_i \cdot \sqrt{{ST}_i} \cdot \sin(2\pi P_i) \\
 {v^z}_i &= |v|_i \cdot \sqrt{1-{{ST}_i}}
\end{align}
where ${ST}_i$ and $P_i$ are uniformly-distributed random variables in $(0,1)$ and
\begin{equation}
 |v|_i = \sqrt{\frac{k_B T}{2m_e}}\sqrt{{{VX}_i}^2+{{VY}_i}^2+{{VZ}_i}^2}
\end{equation}
where ${VX}_i$, ${VY}_i$ and ${VZ}_i$ are standard normal distributed random variables.

\begin{table*} [ht!]
  \begin{tabular}
    {| c || c || c |}
    \hline
    Variable Name & Typical Value & Description \\
    \hline
    \hline
    WaistMultiplierY & 0.85 & correction factor for measured beam size\\
    WaistMultiplierZ & 0.85 & correction factor for measured beam size\\
    PulseEnergy & 6.0 mJ & laser pulse energy \\
    PulseFWHM & 5.5 ns & FWHM of temporal pulse shape \\
    LaserTimeOffset & 45 ns & mean difference of Ps emission to laser time \\
    MisalignmentY & -1.2 mrad & angle between laser wave-fronts (XY-plane) \\
    MisalignmentZ & 0 mrad & angle between laser wave-fronts (XZ-plane) \\
    PositroniumTemp & 550 K & temperature for velocity distribution \\
    DistanceSourceLaserX & 150 mm & X-axis distance target / mirror \\
    DistanceSourceLaserY & 0 mm & Y-axis distance target / laser \\
    DistanceSourceLaserZ & 3 mm & Z-axis distance target / laser \\
    SigmaBeamX & 1.5 mm & X-axis spread of Ps creation area \\
    SigmaBeamY & 2.0 mm & Y-axis spread of Ps creation area \\
    SigmaEmission & 40 ns & Ps emission time width\\
    DistanceMCPLaserX & 0 mm & X-axis distance laser / MCP \\
    DistanceMCPLaserY & -32.1 mm & Y-axis distance laser / MCP \\
    DistanceMCPLaserZ & 26.8 mm & Z-axis distance laser / MCP \\
    TargetAngle & $45{}^\circ$ & target angle relative to positron beam \\
    \hline
   \end{tabular} 
\caption{Experimental variables and typical values used in Monte-Carlo simulation (run 2)
\label{tab:lasersimvars_run2}}
\end{table*}

The relativistic Doppler shift (up to second order) for this specific path is then calculated using the velocity vector, MisalignmentY ($\delta_y$) and MisalignmentZ ($\delta_z$), which represent the misalignment angles with respect to the laser propagation axis along Y and Z. Since $\delta_y \ll 1$ and $\delta_z \ll 1$ it is given simply by
\begin{equation}
 \Delta {\nu^{DS}}_i = \nu_0 \left(1 + \frac{{v^y}_i \delta_y + {v^z}_i \delta_z}{2c} - \frac{{|v|_i}^2}{2c^2} \right) 
\label{eq:2ndDS}
\end{equation}
where $\nu_0$ is the unperturbed transition frequency.

Since the atoms can only interact with the laser when passing the laser aperture of $6\,\mathrm{mm}$ diameter (see figure \ref{fig:Run1_MCPs} and \ref{fig:Run2_MCPs}), the state populations in the 1S and 2S will only be calculated while they pass this region (see below). Otherwise they will be tracked and the populations corrected according to their decay rates. Positronium, which is photo-ionized, is assumed to be detected regardless of the path, since a small bias voltage on the target transports it towards the MCP. Within reasonable limits for the velocity of the positronium atoms, the pulse energy used for delayed photo-ionization is sufficiently large, such that the ionization rate is close to unity. Likewise, the pulse energy used for the 2S-20P excitation is expected to saturate the transition. Therefore these effects do not need be be modeled separately in the simulation.

Excited states, which are detected by field-ionization, are taken into account only if their path intersects with the active area of the MCP. For these states an additional correction for losses on passing the two grids of
\begin{equation}
 \hat\rho_i = \rho_i \cdot \mathrm{OA}^2 \cdot \left( \vec v_i\cdot \vec e_g \right)^2
\end{equation}
is taken into account, where $\rho_i$ are the state populations (see equation \ref{eq:density-diffeq}), $\mathrm{OA}=92\,\%$ is the open area of the used mesh and $\vec e_g$ is the surface normal vector of the grid. 

\begin{table*} [ht!]
  \begin{tabular}
    {| c || c || c |}
    \hline
    Variable Name & Unit & Description \\
    \hline
    \hline
    LaserFrequency & MHz & excitation detuning (see eq. \ref{eq:detuning}) \\
    PIProb & \% & detection probability for direct PI on the MCP (mean)\\
    Pop1S & \% & fraction of ground state atoms reaching the MCP (mean)\\
    Pop2S & \% & fraction of excited state atoms reaching the MCP (mean)\\
    HitTime & ns & Histogram of 2S atoms arrival time on the MCP \\
    HitTheta & rad & Histogram of 2S atoms angle reaching the MCP \\
    HitPop2S & \% & Histogram of 2S atoms fraction reaching the MCP \\
    Velocity & m/s & Histogram of 2S atoms velocity reaching the MCP \\
    \hline
   \end{tabular} 
\caption{Output variables of Monte-Carlo simulation
\label{tab:lasersimoutput}}
\end{table*}

For each atom, the laser pulse energy and time spread is varied according to a Gaussian distribution, where the standard deviation was taken from data. For run 1 the spread was found to be 3.11\% for the pulse energy and 4.7\% for the time spread, while for run 2 the respective values were measured to be 3.68\% and 3.44\%.

The evolution of state populations are calculated by numerical integration of the optical Bloch equations (see equation \ref{eq:density-diffeq}) along the propagation path within the laser aperture. Before each integration step, the laser intensity seen by the atom at that moment (see equations \ref{eq:pulsedpowerone} and \ref{eq:pulsedpowertwo}), is calculated according to the spatial laser profile and the current time. Then the intensity dependent quantities governing the state evolution are determined, which includes the excitation rate (equation \ref{eq:rabifreq}), ionization rate (equation \ref{eq:ionirate}) and AC Stark shift (equation \ref{eq:starkshift}). Finally, the excitation detuning (equation \ref{eq:detuning}) is calculated including the current chirped laser detuning (corrected by the intensity-weighted average chirp), relativistic Doppler and AC stark shifts and the Bloch equations are integrated. This is continued until the atoms leave the region defined by the laser aperture on the target holder.

This process is repeated until the required number of events and laser detuning steps is reached. The simulation saves the results of each individual atom for later analysis. A list of the available output variables can be found in table \ref{tab:lasersimoutput}.

\subsection{Simulation results\label{subsec:simlaser}}

In this section we present the results of the simulation in general terms, what is the impact of different effects on the line shape, and the fitting procedures employed. Unless noted otherwise, the data shown corresponds to the best fit obtained for the analysis of run 1. Unless otherwise stated, frequency values given in this section usually refer to detuning in the excitation laser, which corresponds to half the value in the transition, since the excitation is a two-photon process.

Since the transition is determined both by direct photo-ionization in the exciting laser and by delayed ionization, two distinctly different line shapes are measured. Generally, direct photo-ionization produces a significantly broader line since it effectively limits the probed atoms lifetimes. Furthermore, the profile is generally more skewed, since the process strongly favors higher laser intensities (see also discussion of the AC Stark shift below). 

The natural line width of the 1S-2S transition is dominated by the annihilation lifetimes and calculation yields
\begin{equation}
\Delta \omega_n \approx \gamma_{1s} + \gamma_{2s} \approx 2\pi \cdot 1.3 \, \mathrm{MHz}
\text{ .}
\label{eq:naturallinewidth}
\end{equation}
However, broadening of the line shape to at least $\Delta \nu \approx 100\,\mathrm{MHz}$ in the transition is evident in all line shapes presented in this work. The main effects responsible for this and other features of the line shape will be explained in the following.

A major source of broadening is the limited interaction time of the positronium atoms with the light field, either due to spatial or temporal properties of the laser beam. For Gaussian profiles the time-bandwidth product reads
\begin{equation}
 \Delta \nu \cdot \Delta t \approx 0.44 
\end{equation}
which is a consequence of the uncertainty principle. To give a numerical example, when a stationary atom interacts with a pulsed laser with Gaussian temporal profile and $5\,\mathrm{ns}$ FWHM, this equates to
\begin{equation}
 \Delta \nu \approx \frac{0.44}{5\,\mathrm{ns}} = 88\,\mathrm{MHz}
\end{equation}
broadening or equivalently to laser bandwidth. If the interaction time is limited by the atom propagating through the laser, this effect is commonly referred to as time-of-flight or transit-time broadening. 

\begin{figure}
\includegraphics[width=\columnwidth]{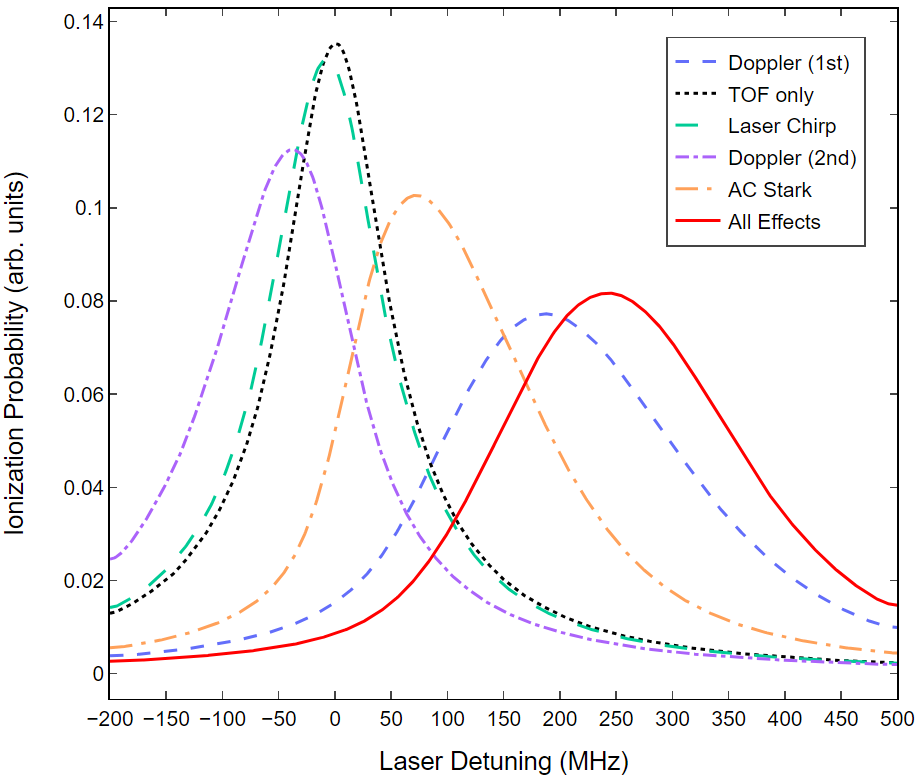} 
   \caption[Impact on the direct photo-ionization line shape for effects included in the Monte-Carlo simulation]{Impact on the direct photo-ionization line shape for effects included in the Monte-Carlo simulation shown for the parameters found as best fit to the data acquired during run 1.  
\label{fig:simulation-effects-pi}}
\end{figure}

\begin{figure}
\includegraphics[width=\columnwidth]{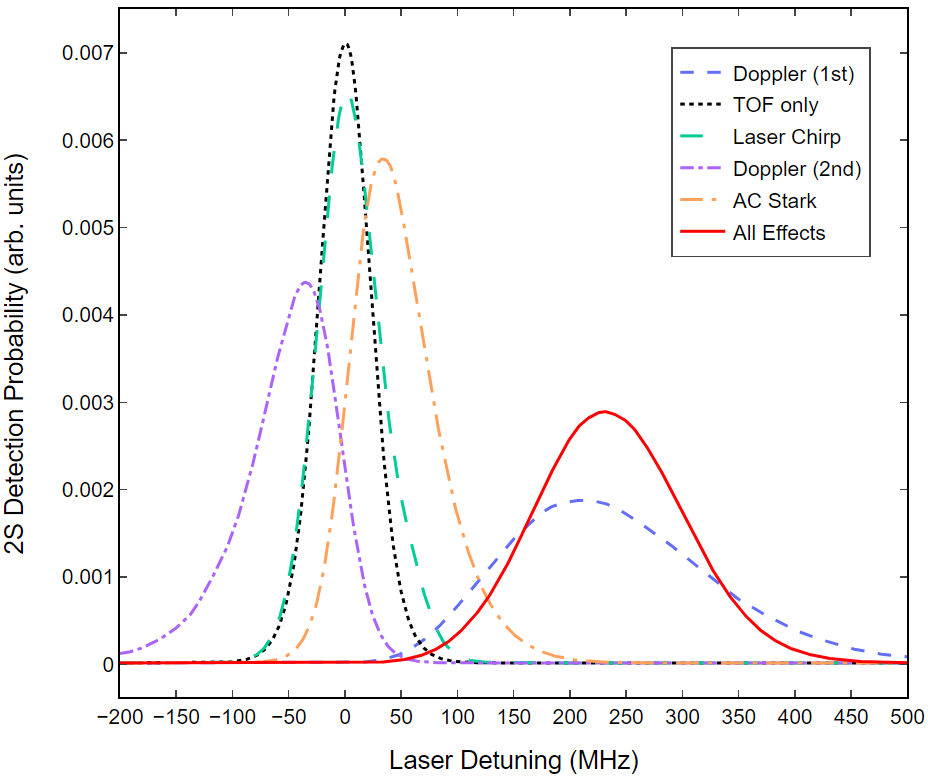} 
   \caption[Impact on the excitation line shape for effects included in the Monte-Carlo simulation]{Impact on the excitation line shape for effects included in the Monte-Carlo simulation shown for the parameters found as best fit to the data acquired during run 1.
\label{fig:simulation-effects-2s}}
\end{figure}

Figures \ref{fig:simulation-effects-pi} and \ref{fig:simulation-effects-2s} shows the impact of other effects included in the simulation on the photo-ionization and excitation line shapes. The graphs correspond to the best fit set of parameters determined for run 1. Run 2 shows analogous behavior unless otherwise noted. Different effects do not simply add up but can have a somewhat complex interaction. Nevertheless, the general effects on the line shape can be identified by studying their impact separately. To this end the plots show the strictly symmetric line shape, which includes TOF broadening only, in black as a starting point. In addition to the impact of the relevant effects included in the simulation shown separately, the final line shape with all effects combined is plotted.

Laser chirp introduces a slight shift and skew of the line shapes. Notably, it causes a red-shift for photo-ionization and a blue-shift for the excitation line, which can be explained by the fact that the chirp is larger for the high intensity part of the laser pulse (see figure \ref{fig:chirp-sim}). Excitation rates generally saturate significantly sooner at higher intensity than photo-ionization rates, which is why intensity-dependent effects generally affect the line-shapes differently.

\begin{figure}
\includegraphics[width=\columnwidth]{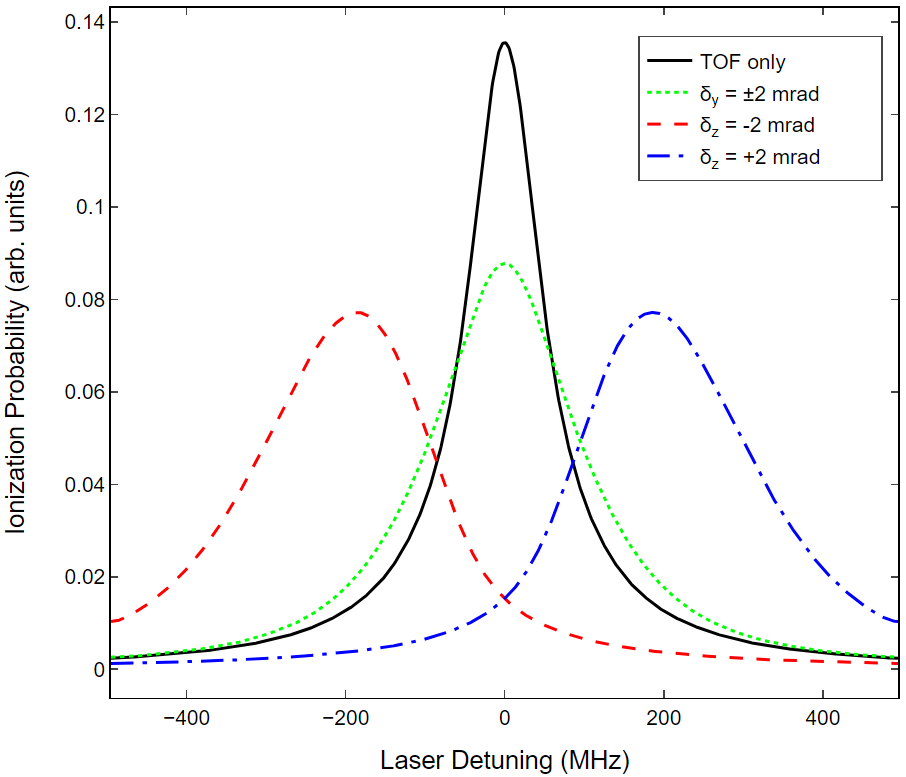} 
   \caption[Impact of the residual first order Doppler shift for different magnitude and direction of misalignment on the direct photo-ionization line shape]{Impact of the residual first order Doppler shift for different magnitude and direction of misalignment on the direct photo-ionization line shape shown for the parameters found as best fit to the data acquired during run 1. 
\label{fig:rfds-pi}}
\end{figure}

The AC Stark effect always blue-shifts, broadens and positively skews the line. Similarly to chirp, the effect is more pronounced in photo-ionization, since it favors higher laser intensities, which makes it one of the major sources of broadening. The effects from second order Doppler shift are of similar magnitude, but opposite sign. While mono-energetic positronium would lead to a constant shift, the distribution of velocities leads to the broadening and skew of the line.

\begin{figure}
\includegraphics[width=\columnwidth]{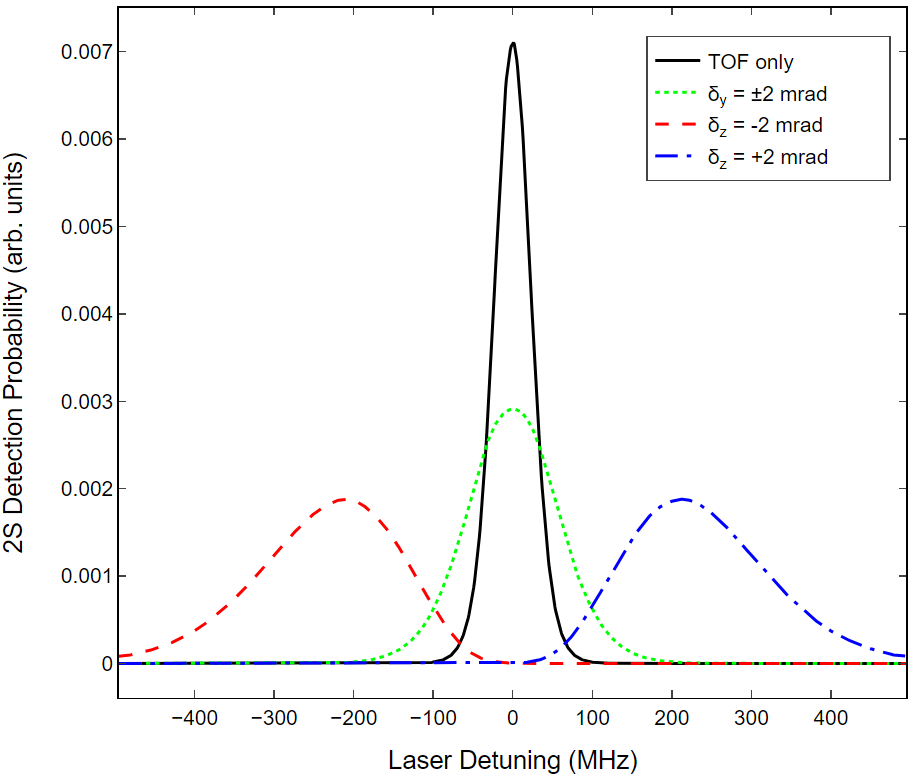} 
   \caption[Impact of the residual first order Doppler shift for different magnitude and direction of misalignment on the excitation line shape (run 1)]{Impact of the residual first order Doppler shift for different magnitude and direction of misalignment on the excitation line shape shown for the parameters found as best fit to the data acquired during run 1.   
\label{fig:rfds-2s}}
\end{figure}

\begin{figure}
\includegraphics[width=\columnwidth]{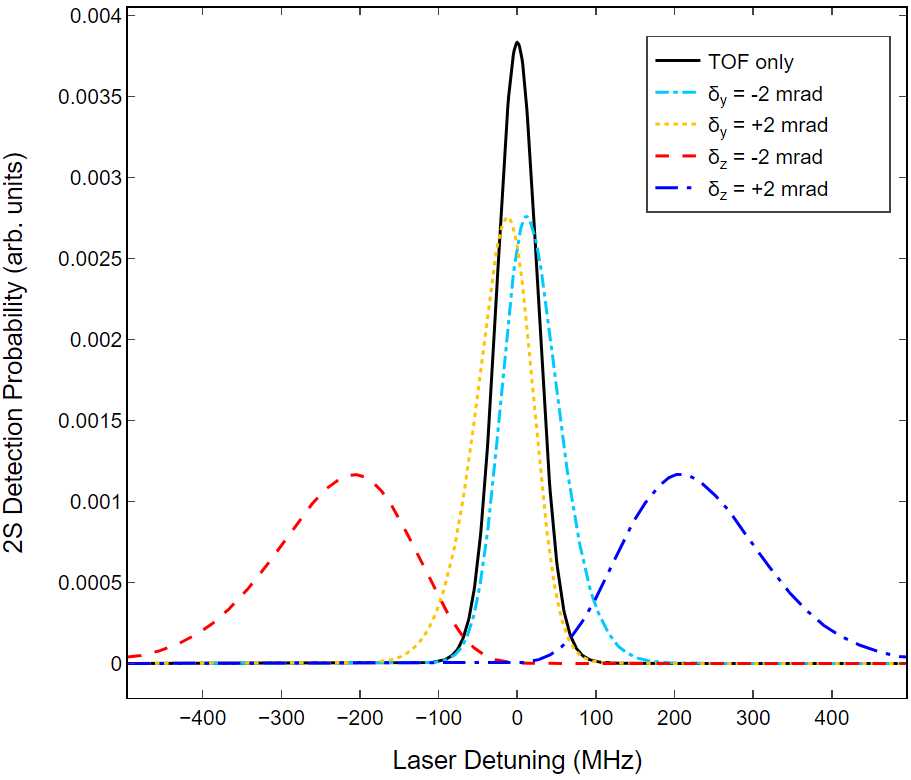} 
   \caption[Impact of the residual first order Doppler shift for different magnitude and direction of misalignment on the excitation line shape (run 2)]{Impact of the residual first order Doppler shift for different magnitude and direction of misalignment on the excitation line shape shown for the geometry used during run 2.   
\label{fig:rfds-2s-run2}}
\end{figure}

The impact of the residual first order Doppler shift is considerably more complex. Generally, misalignment of the laser axis parallel to the target leads to broadening, but not shift (see exception below). The reason for this is that the $\cos \theta$-distribution of emission is axially symmetric, so for every blue-shifted velocity vector, there is an equally likely red-shifted velocity vector. However, this cancellation does not occur for misalignment perpendicular to the target, since all of the possible directions of emitted positronium have at least some velocity component in this direction. This leads to shifts which can have significant magnitude and arbitrary signs, depending on the direction of the misalignment. The resulting shifts for a misalignment of $2\,\mathrm{mrad}$ for both signs along both axes can be found in figures \ref{fig:rfds-pi} and \ref{fig:rfds-2s}. The shifts are qualitatively equivalent in both the photo-ionization and the excitation line shapes. However, the width of broadening in the XY-plane is reduced for the excitation line shape, depending on the angular acceptance of atoms undergoing field-ionization on the MCP. It is important to note that the symmetry of shifts for misalignments in the XY-plane is not just a consequence of the properties of the emission, but also of the detection. In fact, in the case of the experimental geometry used for run 2 this symmetry is broken, due to the angle and offset of the MCP relative to the target. This leads to a small, but significant shift in addition to the broadening in the excitation lineshape, as can be seen in figure \ref{fig:rfds-2s-run2}. The reduced broadening due to smaller angular acceptance of the MCP, as described above, is also evident in this plot.

Ideally, each line would be simulated with very high statistics ($>10^5$) for a very fine scan of laser detunings ($<1\,\mathrm{MHz}$) over a wide range ($\approx 1\,\mathrm{GHz}$) to directly fit measurement data. However, while this might be feasible for a handful of parameters, it is clearly out of the scope for several thousand sets of parameters, as used for the data analysis presented in the next section. To severely reduce the required amount of computational resources, a fitting procedure was developed to convert simulation results to functional form. This allows for the fitting of data with arbitrary offsets, without further interpolation. While the fits give excellent agreement with the simulated line shapes, this introduces an error on the simulated line center, which has to be added to the experimental results.

To accurately model the features of the line shape an appropriate function is required. While the transition itself would present as a Lorentzian with a width corresponding to the natural linewidth, the effects discussed above change the shape significantly. If the broadening introduced is symmetric and due to a parameter with Gaussian distribution (e.g. for time-of-flight broadening due to a Gaussian spatial profile), the appropriate function to fit would be a Voigt profile, which is given by the convolution of a Lorentzian and a Gaussian profile. However, most effects also introduce skewness to the line shape, which has to be taken into account. Several skewed variations of such profiles exist in the literature, however, most do not capture all relevant features. We found that using an extended approach based on efforts to model infrared absorption peaks \cite{Stancik2008} yielded excellent results for a wide range of parameters. Based on the Voigt profile
\begin{equation}
 V(\nu, \sigma,\gamma) = \int_{-\infty}^\infty G(\nu',\sigma)L(\nu-\nu',\gamma)\, d\nu'
\end{equation}
where
\begin{equation}
 G(\nu, \sigma) = \frac{e^{-\nu^2/(2\sigma^2)}}{\sigma \sqrt{2\pi}}
\end{equation}
is the Gaussian and
\begin{equation}
L(\nu,\gamma) = \frac{\gamma}{\pi(\mu^2+\gamma^2)}
\end{equation}
is the Lorentzian distribution, a frequency dependence is included by sigmoidally varying the width parameters, such that
\begin{equation}
\sigma(\nu, \alpha) = \frac{2\sigma_0}{1+e^{\alpha \nu}}
\end{equation}
and analogously
\begin{equation}
\gamma(\nu, \beta) = \frac{2\gamma_0}{1+e^{\beta \nu}}
\end{equation}
where $\alpha$ and $\beta$ are the skewness parameters.

\begin{figure}
\includegraphics[width=\columnwidth]{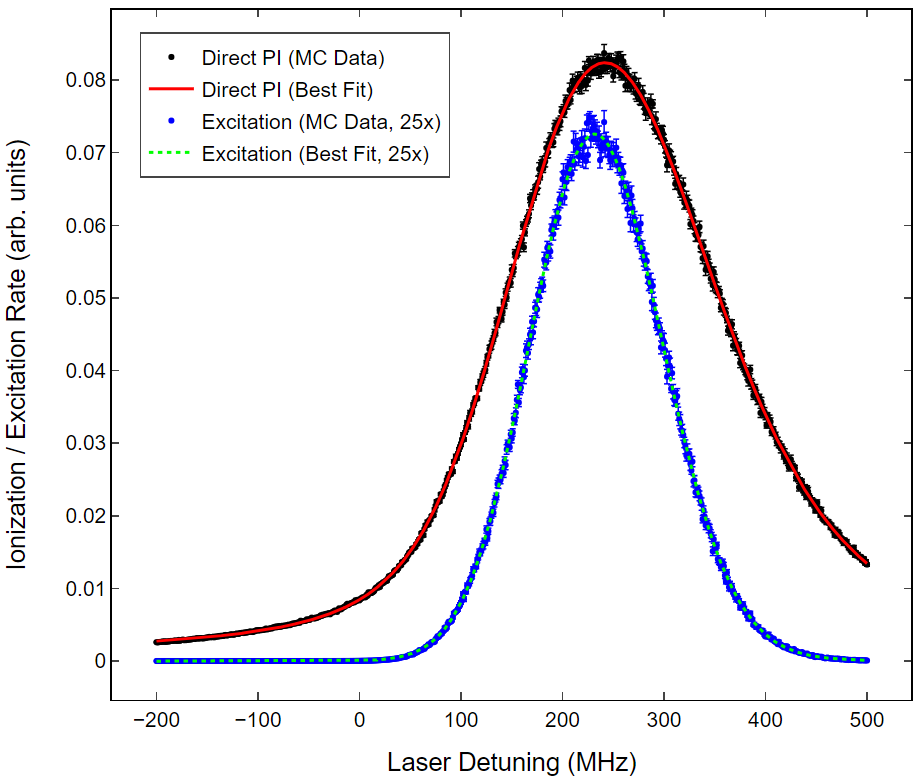} 
   \caption[High statistics line shape simulation for direct photo-ionization and excitation, including best fits.]{High statistics line shape simulation for direct photo-ionization (black) and excitation (blue/grey), including best fits, using doubly skewed Voigt profiles for the most favoured parameter set during run 1.   
\label{fig:sim-fit-line}}
\end{figure}

Figure \ref{fig:sim-fit-line} shows the output and corresponding fits of a high statistics simulation for the best fit parameter set for run 1. The fits have the form
\begin{equation}
\begin{aligned}
 LS(\nu,\nu_{\text{loc}},\sigma_0,\gamma_0,\alpha,\beta,A) = \\ A \cdot V((\nu-\nu_{\text{loc}}), \sigma(\alpha),\gamma(\beta))
 \end{aligned}
\end{equation}
where $\nu_{\text{loc}}$ is a location parameter, that corresponds to the mean value in the limit of no skewness and $A$ is an overall amplitude scaling parameter. The set of parameters of the best fit shown in figure \ref{fig:sim-fit-line} of the photo-ionization line is given by
\begin{align}
 &{\nu_{\text{loc}}}^{\text{PI}} = (247.98 \pm 0.14)\,\mathrm{MHz} \\
 &\sigma^{\text{PI}} = (80.0 \pm 0.53)\,\mathrm{MHz} \\
 &\gamma^{\text{PI}} = (54.78 \pm 0.82)\,\mathrm{MHz} \\
 &\alpha^{\text{PI}} = -2.564 \times 10^{-3} \pm 5.3\times 10^{-5} \\
 &\beta^{\text{PI}}  = -6.2 \times 10^{-5} \pm 1.2 \times 10^{-4} \\
 &A^{\text{PI}} = 26.431 \pm 0.066
\end{align}
while the parameters for the best fit of the excitation lineshape read
\begin{align}
 &{\nu_{\text{loc}}}^{\text{2S}} = (232.92 \pm 0.10)\,\mathrm{MHz} \\
 &\sigma^{\text{2S}} = (65.41 \pm 0.22)\,\mathrm{MHz} \\
 &\gamma^{\text{2S}} = (0.07 \pm 0.32)\,\mathrm{MHz} \\
 &\alpha^{\text{2S}} = -6.11 \times 10^{-4} \pm 6.5\times 10^{-5} \\
 &\beta^{\text{2S}}  = -0.01 \pm 0.10 \\
 &A^{\text{2S}} = 0.47604 \pm 8.7 \times 10^{-4}
\end{align}
where the error on the location parameter $\nu_{\text{loc}}$ corresponds approximately to the error of the line center (since $\alpha \ll 1$ and $\beta \ll 1$).

\begin{figure}
\includegraphics[width=\columnwidth]{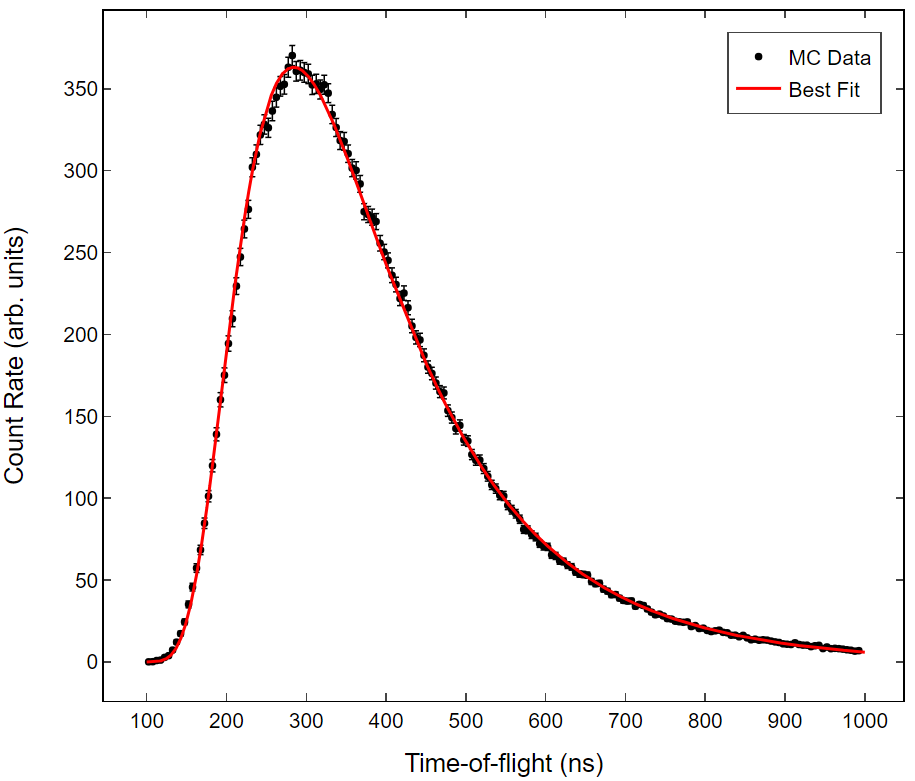} 
   \caption[High statistics time-of-flight spectrum simulation, including best fit.]{High statistics time-of-flight spectrum simulation (black), including best fit (red) using a $\chi$-distribution for the most favored parameter set of run 1.   
\label{fig:sim-fit-tof}}
\end{figure}

In addition to the line shape, the time-of-flight distribution is fitted to allow for direct comparison with measurement. It has been found that a $\chi$-distribution
\begin{equation}
 \chi(x, k) = \frac{x^{k-1}e^{-\frac{x^2}{2}}}{2^{\frac{k}{2}-1} \Gamma\left(\frac{k}{2}\right)}
\end{equation}
with arbitrary (continuous) degrees of freedom $k$ fits the spectra exceptionally well. Figure \ref{fig:sim-fit-tof} shows the TOF spectrum corresponding to the line shapes of figure \ref{fig:sim-fit-line} using the function
\begin{equation}
 \text{TOF}(t, l, s, k, B) = B \cdot \chi\left(\frac{\frac{1}{t} - l}{s}, k\right)
\end{equation}
where the parameter $l$ gives location, $s$  time scaling and $B$ overall amplitude. The best fit parameters were found to be
\begin{align}
 &l = (5.31 \times 10^{-4} \pm 1.3 \times 10^{-5})\,\frac{1}{\mathrm{s}} \\
 &s = (1.7231 \times 10^{-3} \pm 4.7 \times 10^{-6})\,\frac{1}{\mathrm{s}} \\
 &k = 4.00 \pm 0.04 \\
 &B = 1079.5 \pm 3.0 \quad \text{.}
\end{align}
Note that the choice of distribution is motivated by the Maxwell-Boltzmann velocity profile, which corresponds to a $\chi$-distribution with $k=3$. However, the parameters do not have a direct physical correspondence since the distribution is a convolution of velocities, excitation and angular emission probabilities.

Several thousand combinations of parameters within experimental limits were used for the Monte-Carlo simulations to cover all possible parameter space for both experimental runs. Each line-shape was then fitted according to the description above, which provides a database of functions to fit experimental results, extract the transition frequency, and estimate the systematic errors of the measurement.

\section{\label{sec:appendix-ProspectCW}Prospects for a CW measurement}

\begin{figure}
\includegraphics[width=0.968\columnwidth]{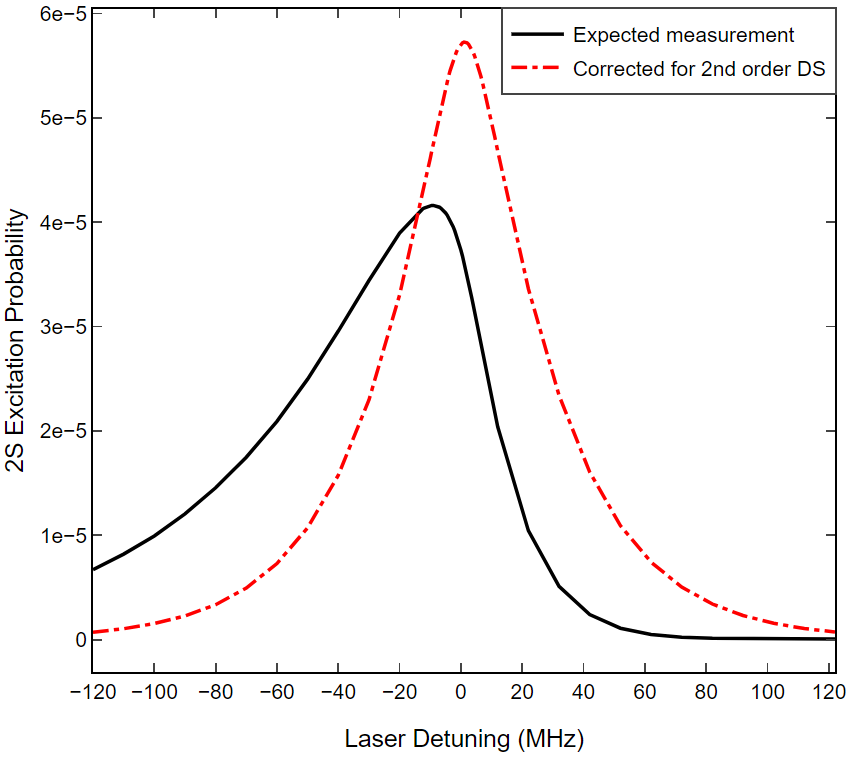} 
   \caption{Predicted lineshape for CW excitation at \SI{450}{\watt} and \SI{700}{\kelvin} velocity distribution before (black solid) and after (red dashdotted) the subtraction of the second order Doppler shift.    
\label{fig:sim-CW-line}}
\end{figure}

To demonstrate how the method presented in this work enables achieving sub-MHz uncertainty in a future measurement using a CW laser, we present in Figure \ref{fig:sim-CW-line} the simulated lineshape before and after applying the second-order Doppler shift correction under realistic experimental conditions with a CW source.

The circulating power is 450 W with laser (310 micron beam waist) positioned 3.5 mm from the positronium formation target. Under these conditions, the AC Stark effect induces a shift of approximately 2 MHz on the atomic transition.

For the runtime estimation, we assume the use of a new standard 1 GBq positron source, a factor 10 times more intense than the source utilized in this study. With a 1 GBq source, about  $10^5$ positrons per pulse are expected. With our typical 20\% positronium conversion efficiency,  50\% Rydberg 20P excitation probability, 50\% detection efficiency and considering the losses due to atoms decaying from the excitation region to the MCP placed in 10 cm distance, we estimate that a 1 MHz determination is possible in 3 days of runtime. 

To improve the velocity reconstruction accuracy, a position-sensitive MCP can be employed. Assuming a spatial resolution of 100 microns, a time resolution of $\sigma_t=10\,\mathrm{ns}$ (accounting for the positron bunch length, positronium diffusion time, and MCP resolution), and a $\sigma_e=2\,\mathrm{mm}$ initial spatial emission profile, the error in reconstructing the atoms’ velocities (around 1\%) leads to a systematic uncertainty of approximately 50 kHz estimated using Equation \ref{eq:2ndDS}.

The second largest contribution to the systematic is the AC Stark shift, which can be controlled by stabilizing the circulating power in the cavity to within 1\%, resulting in an uncertainty of 20 kHz. This can be further cross-checked by measuring at three different power levels and extrapolating to zero power.

The frequency chirp observed in pulsed laser systems will not be present in this experiment, since a CW laser source will be used. The uncertainty associated with the frequency comb, referenced to a GPS as in run 2, is at the kHz level. Additionally, as no magnetic fields are used to guide positrons and no electric fields are applied, the Zeeman and DC Stark shifts are expected to contribute negligibly (below 1 kHz).

\bibliographystyle{unsrt}
\bibliography{literature}

\end{document}